\documentclass[fleqn,usenatbib]{mnras}

\usepackage[T1]{fontenc}
\usepackage{ae,aecompl}

\usepackage{graphicx}	% Including figure files
\usepackage{amsmath}	% Advanced maths commands
\usepackage{amssymb}	% Extra maths symbols
\usepackage[dvipsnames]{xcolor} % highlighting
\usepackage{cuted}
\usepackage{xspace}

\newcommand{\msun}{{M}_\odot}
\newcommand{\mste}{M_{\rm *}}
\newcommand{\mvir}{M_{\rm vir, peak}}
\newcommand{\asloth}{\textsc{a-sloth}\xspace}
\newcommand{\ctp}{\textit{Caterpillar}\xspace}
\title[]{Tracing stars in Milky Way satellites with A-SLOTH}
\author[Li-Hsin Chen]{
Li-Hsin Chen$^{1,2}$, Mattis Magg$^{1,2}$, Tilman Hartwig$^{3,4,5}$, Simon C. O. Glover$^{1}$,
\newauthor
~Alexander P. Ji$^{7,8}$ and Ralf S. Klessen$^{1,6}$ 
\\
$^{1}$Universit\"{a}t Heidelberg, Zentrum f\"{u}r Astronomie, Institut f\"{u}r Theoretische Astrophysik, Albert-Ueberle-Str.\ 2, 69120 Heidelberg, Germany\\
$^{2}$International Max Planck Research School for Astronomy and Cosmic Physics at the University of Heidelberg (IMPRS-HD), \\ Königstuhl 17, D-69117 Heidelberg, Germany \\
$^{3}${Department of Physics, School of Science, The University of Tokyo, Bunkyo, Tokyo 113-0033, Japan}\\
$^{4}${Institute for Physics of Intelligence, School of Science, The University of Tokyo, Bunkyo, Tokyo 113-0033, Japan}\\
$^{5}$Kavli Institute for the Physics and Mathematics of the Universe (WPI), The University of Tokyo Institutes for Advanced Study, \\ The University of Tokyo, Kashiwa, Chiba, 277-8583, Japan\\
$^{6}$Universit\"{a}t Heidelberg, Interdisziplin\"{a}res Zentrum f\"{u}r Wissenschaftliches Rechnen, Im Neuenheimer Feld 225, 69120 Heidelberg, Germany \\
$^{7}$Department of Astronomy \& Astrophysics, University of Chicago, 5640 S Ellis Avenue, Chicago, IL 60637, USA \\
$^{8}$Kavli Institute for Cosmological Physics, University of Chicago, Chicago, IL 60637, USA % ORCID 0000-0002-4863-8842
}   
\date{}

\begin{document}
\label{firstpage}
\pagerange{\pageref{firstpage}--\pageref{lastpage}}
\maketitle

\begin{abstract}
We study the stellar mass-to-halo mass relation at $z=0$ in 30 Milky Way-like systems down to the ultra-faint ($\mste<10^5\msun$) regime using the semi-analytic model \asloth. A new model allows us to follow star formation and the stochastic stellar feedback from individually sampled Pop II stars. Our fiducial model produces consistent results with the stellar mass-to-halo mass relation derived from abundance matching and the observed cumulative stellar mass function above the observational completeness. We find a plateau in the stellar mass-to-halo mass relation in the ultra-faint regime. The stellar mass of this plateau tells us how many stars formed before supernovae occur and regulate further star formation, which is determined by the Pop~II star formation efficiency. We also find that the number of luminous satellites increases rapidly as $\mste$ decreases until $\mste \approx 10^4\msun$. Finally, we find that the relative streaming velocity between baryons and dark matter at high redshift is important in determining the number of ultra-faint dwarf galaxies at $z=0$. The new model in \asloth provides a framework to study the stellar properties and the formation history of metal-poor stars in Milky Way and its satellites.
\end{abstract}


\section{Introduction}
\label{sec:intro}
Galaxy formation depends heavily on properties and evolution of the host dark matter halo. The most straightforward connection between them is perhaps the stellar mass-to-halo mass (SMHM) relation. Previous studies have found that for systems with stellar masses $\mste > 10^5\msun$ \citep{Shankar:2006aa,Drlica-Wagner:2020aa, Garrison-Kimmel:2014ab,Garrison-Kimmel:2017aa,Jethwa:2018aa,Salucci:2019aa,Nadler:2020aa}, the galaxies and their host haloes follow a tight SMHM relation. However, whether this relation still holds for ultra-faint dwarf galaxies ($\mste< 10^5\msun$, UFDs) is still unclear. In recent years, numerous UFDs around the Milky Way (MW) have been discovered by large surveys \citep{Willman:2010aa,Drlica-Wagner:2015aa,Drlica-Wagner:2020aa,Koposov:2015aa,Torrealba:2016aa,Torrealba:2018aa}. 
Despite the low completeness of UFD discoveries, these UFDs already provide us with constraints on the SMHM relation and the underlying galaxy formation model.

There have been many recent cosmological zoom-in simulations of MW-like systems or isolated UFDs \citep[e.g.][]{Hopkins:2014aa,Wetzel:2016aa,Wheeler:2019aa,Libeskind:2020aa,Font:2020aa,Font:2021aa,Engler:2021aa}.
These simulations still cannot resolve the smallest dwarf galaxies in the MW system. Due to the high spatial and mass resolutions that are needed to properly simulate UFDs, the required time and computational resource is immense.
Semi-analytic models provide an opportunity to explore different physical processes and a wider range of parameters. 
For example, \citet{deBennassuti:2014aa,deBennassuti:2017aa} utilized the semi-analytic code \textsc{gamete} to investigate the metallicity distribution function in the Galactic halo and the carbon-enhanced metal-poor stars. Their model is based on the Extended Press–Schechter formalism \citep{Press:1974aa,Lacey:1993aa} and includes the transition between metal-free and metal-enriched star formation following metal and dust evolution. 
\citet{Salvadori:2015aa} and \citet{Rossi:2021aa} employed \textsc{gamete} to study the carbon-enhanced metal-poor stars in the dwarf galaxies in the Local Group. 
\citet{Visbal:2018aa,Visbal:2020aa} studied the metal-free star formation rate down to $z=6$, including physical processes such as photo-ionization, Lyman-Werner (LW) radiation, and metal enrichment.
\citet{Kravtsov:2021aa} used \textsc{grumpy} to study the stellar mass-to-halo mass relation and stellar mass-metallicity relations in the dwarf galaxies. However, they do not include metal-free star formation in their model. 
% \lhnote{Note that this is not a very complete list; you should perhaps specify that you focus just on things that include Pop III stars.}

In this work, we use our semi-analytic galaxy formation code \asloth (Ancient Stars and Local Observables by Tracing haloes, \textcolor{blue}{Hartwig et al. in prep)} to study the SMHM relation in satellites of MW-like systems. 
\asloth has been used to study various problems, such as the lower limit of initial mass function (IMF) of metal-free stars \citep{Hartwig:2015aa}, the connection between the metal-poor stars in the MW and their progenitors \citep{Hartwig:2018aa, Hartwig:2019aa}, the probability of finding metal-free survivors in the MW \citep{Magg:2018aa}, and the inhomogeneous mixing of metals in the interstellar medium \citep{Tarumi:2020ab}. 
In these previous studies, \asloth only tracked the total mass of metal-enriched stars forming in each system, rather than tracing the formation and evolution of the stars individually. Therefore, the corresponding stellar feedback was considered to be continuous and computed using IMF-averaged quantities. In this paper, we introduce a new model in \asloth that allows metal-enriched stars to be traced individually. This enables \asloth to properly follow the detailed star formation activity in individual systems and the impact of stellar feedback. 
We describe the general structure and the star formation models of metal-free and metal-enriched stars in Section~\ref{sec:nm}. We describe how we calibrate the free parameters in our model in Section~\ref{sec:cali}. We discuss the main results in Section~\ref{sec:results}. We discuss the implications of our new model and results in Section~\ref{sec:dis}. Finally, we summarize this work in Section~\ref{sec:sum}.

\section{Numerical Method}
\label{sec:nm}
A complete description of the physical processes accounted for in \asloth is given in \textcolor{blue}{Hartwig et al. (in prep)}. Here, we focus on the parts of the model that are the most important for our current study. 

\subsection{Merger trees}
\label{sec:mt}
\asloth uses dark matter halo merger trees drawn from the \textsc{lx}14 run of the \ctp project \citep{Griffen:2016aa}, which is a dark-matter-only cosmological simulation suite focusing on the assembly of MW-sized haloes and their satellite galaxies. The dark matter particles in the \textsc{lx}14 run have masses of $\sim 3 \times 10^4 \msun$ and the gravitational softening length is $76 h^{-1}$pc. It is based on the \citet{Planck13XVI} cosmological parameters in a 100 $h^{-1}$Mpc box. MW-like systems are selected at $z=0$ if the main haloes fulfill the following criteria: 
\begin{enumerate}
  \item \noindent Virial mass in range $0.7 \times 10^{12}\msun$ $\leq \mvir \leq 3 \times 10^{12}\msun$. 
  \item \noindent There is no halo with $\mvir \geq 7 \times 10^{13}\msun$ within 7~Mpc. 
  \item \noindent There are no other haloes with $\mvir \geq 0.5 \times M_{\rm main}$ within 2.8 Mpc of the main halo. 
\end{enumerate}
Dark matter haloes are identified with \textsc{rockstar} \citep{Behroozi:2013aa} and $\mvir$ is the maximum virial mass along the main branch a halo ever attains up to the current redshift, that is computed with the evolution of the virial relation from \citet{Bryan:1998aa}. We adopt their notation here. 
\citet{Griffen:2016aa} extracted the full merger history of these MW-like systems, including dark matter and spatial information of the haloes and global time-steps between the snapshots, $\Delta t_z$, which is $\approx 5$Myr down to $z=6$ and $\approx 50$Myr at $z=6-0$. We have verified in an earlier study that the mass resolution of these merger trees is sufficient to model Pop~III star formation at high redshift and using higher resolution merger trees yields converged results (see Appendix A of \citealt{Magg:2018aa}).
In this work, we select 30 merger trees from their sample. \asloth walks through these merger trees and determines the baryonic contents of each halo in the merger tree based on the implemented physics. We describe the physical processes in the following sections. 

\subsection{Population III (Pop~III) stars}
\label{sec:popiiisf}
We follow the same metal-free (Pop~III) star formation (SF) model as in our previous works \citep{Magg:2018aa,Tarumi:2020ab}. We briefly summarize the model here. Pop~III stars are assumed to form when H$_2$ can cool primordial gas efficiently, which means that a halo needs to exceed a critical mass, $M_{\rm crit}$. In \asloth, Pop~III stars form instantly in a single star burst in a mini-halo (haloes below the atomic cooling threshold ($T_{\rm vir} = 10000\,\mathrm{K}$) as soon as its virial mass exceeds $M_{\rm crit}$. 
By default, we follow the prescriptions given in \citet[][their Eqs.~9 and 10]{Schauer:2021aa} and \citet[][their Eq.~1]{Hummel:2012aa} to determine $M_{\rm crit}$.
\citet{Schauer:2021aa} showed that the critical mass is dependent on the LW background and the large-scale streaming velocity of the baryons relative to the dark matter: 
\begin{equation}
   {\rm log_{10}} M_{\rm crit, S21} = 6.0174 \, ( 1.0 + 0.166 \sqrt{J_{21}} ) + 0.4159 \frac{v_{\rm BC}}{\sigma_{\rm rms}}.
\end{equation}
Here, $M_{\rm crit, S21}$ is the critical mass in units of solar masses, $J_{21}$ is the strength of LW background in units of $10^{-21} \: {\rm erg \, s^{-1} \, cm^{-2} \, Hz^{-1} \, sr^{-1}}$, $v_{\rm BC}$ is the large-scale streaming velocity of the baryons relative to the dark matter in units of $\sigma_{\rm rms}$, and $\sigma_{\rm rms}$ is the root-mean-squared value of the streaming velocity. We denote this default model hereafter as S21. We do not self-consistently follow the build-up of the LW background, because the effective volume of our MW-like merger trees is too small to obtain a cosmologically representative estimate of the LW background. Instead, we adopt the simple redshift-dependent fitting formula,
\begin{equation}
J_{21} = 10^{2-z/5},
\end{equation}
which is based on the work in \citet{Greif:2006aa}. We take $v_{\rm BC} = 0.8\sigma_{\rm rms}$ as the fiducial baryonic streaming velocity, since this is the most likely value to be present at a randomly selected point in the Universe  \citep{Schauer:2021aa}. The impact of varying our treatment of both of these effects is explored in Section~\ref{sec:LWstream}. 

In addition, we assume that gas can cool down efficiently and form stars in haloes that have virial temperatures above the atomic cooling limit ($T_{\rm vir} \geq 10^4{\rm K}$) when there is no strong ionizing radiation field \citep{Visbal:2017aa}. Therefore, we consider haloes that have virial masses larger than $M_{\rm crit, 10^4{\rm K}}$ can form Pop~III stars, where $M_{\rm crit, 10^4{\rm K}}$ is computed by 
\begin{equation}
    M_{\rm crit, 10^4{\rm K}} = 10^{7.5} \left( \frac{1+z}{10} \right)^{-1.5}.
\end{equation}
The final critical mass is then determined by $M_{\rm crit, fin} = {\rm MIN}(M_{\rm crit, S21}, M_{\rm crit, 10^4{\rm K}})$.

Cold gas is converted to Pop~III stars in haloes with $M_{\rm halo} \geq M_{\rm crit}$ at a fixed efficiency $\eta_{\rm III}$, defined such that $M_{{\rm *,III}} = \eta_{\rm III} \Omega_{\rm b} M_{\rm halo} / \Omega_{\rm m} $, where $M_{\rm halo}$ is the current virial mass of the halo, $\Omega_{\rm b}$ is the baryon density parameter today, and $\Omega_{\rm m}$ is the matter density parameter today. We adopt $\Omega_{\rm b} = 0.0486$ and $\Omega_{\rm m} = 0.3089$ \citep{Planck15XIII}. 
We follow the Pop~III model in \citet{Tarumi:2020ab} and sample stars stochastically from an IMF with a slope of $dN/d(\rm{log}M) \propto M^{0.5}$ in the mass range of $2<\mste<180\msun$. 
Further SF is halted when the total mass of newly-formed Pop~III reaches $M_{\rm {*,III}}$. 

We calculate the stellar lifetimes of Pop~III stars by interpolating between values assembled from several previous studies: 
\citet{Marigo:2001aa} for stars in the mass range 0.7-100$\msun$; \citet{Schaerer:2002aa} for stars in the mass range 5-500$\msun$ (and note that we consider their models without mass loss); and \citet{Esktroem:2008aa} for stars in the mass range 9-200$\msun$. If the stellar lifetime of a star is provided by multiple works, we use the mean value. Pop~III stars die as core-collapse supernovae (CCSNe) in the range of 10-40$\msun$ and as pair-instability supernovae (PISNe) in the range of 140-260$\msun$ \citep{Heger:2002aa,Heger:2010aa}. Following \citet{Tarumi:2020ab}, we further assume that $30\%$ ($f_{\rm faint} = 0.3$) of the Pop~III CCSNe explode as faint supernovae. We use the tabulated metal yields provided in \citet{Kobayashi:2011aa} and \citet{Nomoto:2013aa} for PISNe and normal CCSNe, and the yields in \citet{Ishigaki:2014aa,Ishigaki:2018aa} for faint CCSNe. 
We assume that a fraction $f_{\rm fallback} = 0.2$ \citep{Ritter:2015aa} of the metals remains in the mini-halo after the supernovae have exploded, with the remaining fraction $(1 - f_{\rm fallback})$ gaining enough energy from the SNe to escape from the gravitational well of the mini-halo.
\citet{Tarumi:2020ab} calibrated the Pop~III SF efficiency ($\eta_{\rm III}$), $f_{\rm faint}$ and $f_{\rm fallback}$ to reproduce the observed MW metallicity distribution function (MDF). The main focus of this work is on the stellar masses of the dwarf galaxies and our model does not heavily depend on metallicity. Therefore, we adopt the same values of $\eta_{\rm III}$, $f_{\rm faint}$ and $f_{\rm fallback}$ as in \citet{Tarumi:2020ab} without investigating the effects of these parameters in detail.

Following the initial Pop~III star burst, we assume that any future star formation in the same halo will be in the form of metal-enriched stars. However, stellar feedback will have heated the gas in the mini-halo and ejected a fraction of it. Therefore, further SF is suppressed until the gas re-accumulates and cools down. We denote this time between the Pop~III star burst and the onset of further star formation as $t_{\rm rec}$, known as the recovery time, which was calibrated by \citet{Tarumi:2020ab}.

\subsection{Population II (Pop~II) stars} 
\label{sec:popiisf}

\begin{table}
\centering
\begin{tabular}{l|l}
    Name & Definition \\
    \hline
    $\mvir$ & peak virial mass of the halo up to current $z$ \\
    $M_{\rm cold}$ & cold gas mass \\
    $M_{\rm hot}$ & hot gas mass \\
    $M_{\rm out}$ & cumulative outflow mass \\
    $M_{\rm *,II}$ & total Pop II stellar mass \\ 
    $M_{\rm disk}$ & disk mass, including gas and stars \\
    $\delta M_{\rm out}$ & outflow mass \\
    $\delta M_{\rm out, cold}$ & cold gas mass that enters outflow \\
    $\delta M_{\rm out, hot}$ & hot gas mass that enters outflow \\
    $\delta M_{\rm heat}$ & mass that transfers from cold gas to hot gas \\
    $\delta M_{\rm *,II}$ & Pop II stellar mass that is formed \\
    $\delta M_{\rm acc, hot}$ & hot gas mass that is accreted from IGM \\
    $i$ & step $i$ in the subcycle \\
    \hline
    $n_{\rm bins, II}$ & number of Pop~II IMF bins \\
    $M_{\rm bound, II}$ & lower and upper limits of Pop~II stellar mass \\
    $n^{\rm den}_{\rm cold}$ & number density of dense gas \\ 
    $v_{\rm BC}$ & relative streaming velocity between \\ 
    & baryon and dark matter \\
    $\eta_{\rm II}$ & Pop~II star formation efficiency \\
    $\alpha_{\rm II}$ & slope of the Pop~II IMF \\
    $\gamma_{\rm out}$ & outflow efficiency (Eq.~\ref{eq:gammaout})\\
    $\alpha_{\rm out}$ & exponent in $\gamma_{\rm out}$\\
    $M_{\rm norm}$ & normalization mass in $\gamma_{\rm out}$ \\
    \hline
    $\Delta t_z$ & time difference between each snapshot \\
    & in the merger tree \\
    $\delta t_i$ & adaptive time-step $i$ in the SF subcycle \\
    $t_{\rm dyn}$ & dynamical time scale of the halo \\
    $t_{\rm cold, ff}$ & free-fall time scale of cold gas \\
    $t_{\rm star}$ & stellar formation time scale \\
    $t_{\rm cool}$ & cooling time scale \\ 
    $t_{\rm acc}$ & hot gas accretion time scale \\
    \hline
\end{tabular}
\caption{Definitions of the variables in the Pop~II star formation model.
\label{table:notation}}
\end{table}
In our model, we do not distinguish between metal-rich (Population I) and metal-poor (Population II) stars and classify all metal-enriched stars as Population II (Pop~II) stars. 
We implement an improved Pop~II SF model, which is based on the models in \citet{Magg:2018aa} and \citet{Tarumi:2020ab}. The explanations of important variables in the Pop~II SF model are listed in Table~\ref{table:notation}. We assume that the baryons initially associated with a given dark matter halo are located either in cold gas ($M_{\rm cold}$), in hot gas ($M_{\rm hot}$), in a Pop~II stellar component ($M_{\rm *,II}$), or have been lost from the halo in an outflow ($M_{\rm out}$). The initial baryonic mass of a halo with virial mass $\mvir$ is assumed to be $(\Omega_{\rm b} / \Omega_{\rm m})\mvir$. Therefore, by construction
\begin{equation}
\begin{aligned}
    \mvir \frac{\Omega_{\rm b}}{\Omega_{\rm m}} = M_{\rm cold} + M_{\rm hot} + M_{\rm *,II} + M_{\rm out}
\end{aligned}
\end{equation}
is always maintained in the model. Note that in our model, hot gas is any gas that is not cold, i.e.\ it corresponds to the sum of the warm and hot phases of the ISM in the usual three-phase description \citep{McKee:1981aa, Klessen:2016aa}. We do not account for Pop~III stars in the mass budget because the small value we adopt for $\eta_{\rm III}$ means that they never represent more than a small fraction of they baryonic mass in any halo. Additionally, they occupy a negligible fraction of the stellar mass at $z=0$ and the Pop~III stars that remain on the main sequence after $t_{\rm rec}$ are too low mass to contribute significantly to the stellar feedback.
%and the survivors do not provide influential stellar feedback.

Unlike Pop~III SF, we assume that Pop~II stars can form in multiple epochs, and so we further divide the global time-steps ($\Delta t_z$) from the merger tree into smaller ones. In other words, we have subcycles in our Pop~II SF model. The subcycle time-step is denoted as $\delta t_{i}$, and we describe how we determine it in our simulation at the end of this section. A halo is labeled to trigger Pop~II SF if the time since it experienced Pop~III SF equals or exceeds $t_{\rm rec}$ or if it is externally enriched by supernovae from nearby haloes. Prior to the subcycling of Pop~II SF, we initialize the halo by inheriting the baryonic contents from all of its progenitors, which gives us $M^{0}_{\rm cold}$, $M^{0}_{\rm hot}$, $M^{0}_{\rm *,II}$, and $M^{0}_{\rm out}$. Then, in time-step $i$ of the subcycle, the baryonic quantities are updated with the following equations:
\begin{equation}
\begin{aligned}
    & M^{i+1}_{\rm cold} =  M^{i}_{\rm cold} + \frac{\delta t_{i} M^{i}_{\rm hot}}{t_{\rm dyn}} - \delta M^{i}_{\rm out, cold} - \delta M^{i}_{\rm heat} - \delta M^{i}_{\rm *,II}, \\
    & M^{i+1}_{\rm hot} = M^{i}_{\rm hot} - \frac{\delta t_{i} M^{i}_{\rm hot}}{t_{\rm dyn}}  - \delta M^{i}_{\rm out, hot} + \delta M^{i}_{\rm heat} + \delta M^{i}_{\rm acc, hot}, \\
    & M^{i+1}_{\rm out} = M^{i}_{\rm out} + \delta M^{i}_{\rm out, cold} + \delta M^{i}_{\rm out, hot}. \\
\end{aligned}
\end{equation}
We start from $i=0$ and make sure the last time-step in the subcycle always ends exactly at the next global time-step of the merger trees.

There are several physical processes that enter the above equations. We show the connections between these processes and the baryonic contents in our Pop~II SF model in Figure~\ref{fig:fc}: 

\begin{figure}
    \centering
    \includegraphics[width=\columnwidth]{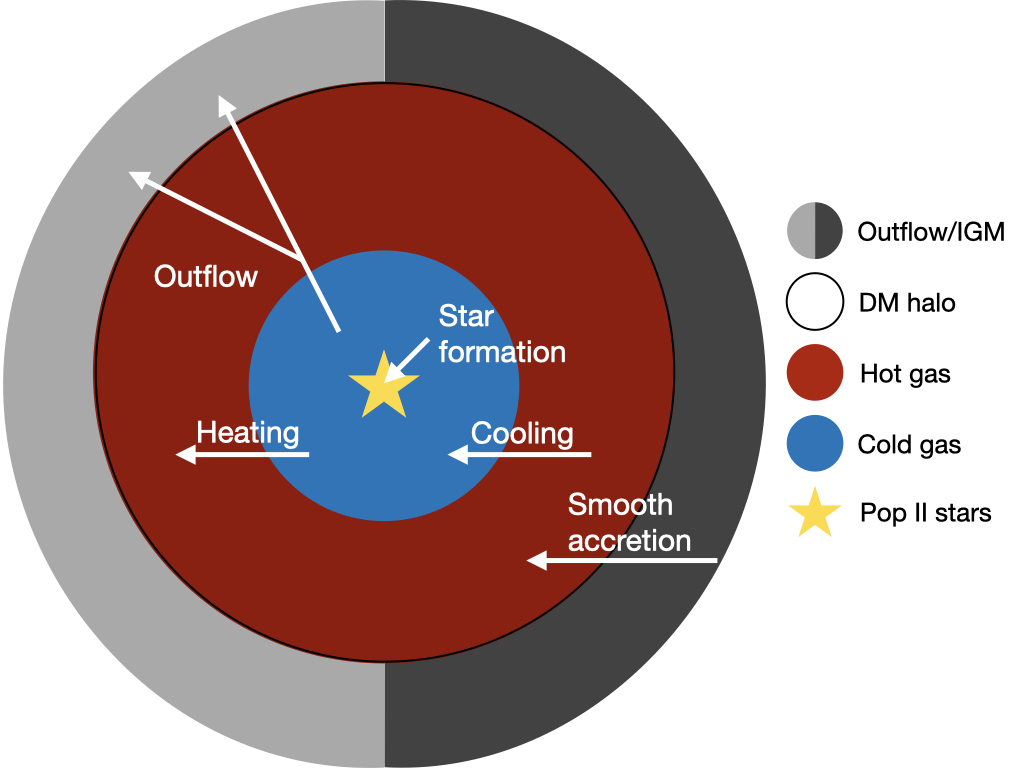}
    \caption{We show the connections between different components of our Pop~II SF model in this cartoon plot. Note that we do not re-accrete outflow material. Therefore we use separate colours for outflow and IGM.}
    \label{fig:fc}
\end{figure}

\begin{enumerate}
  \item {\noindent \bf Cooling of hot gas} 
  
  We assume that hot gas cools on a timescale equal to the dynamical time of the stellar disk, $t_{\rm dyn}$. Therefore, during the subcycle time-step $\delta t_{i}$, a mass of gas $ M^{i}_{\rm hot} \delta t_{i}/{t_{\rm dyn}}$ is transferred from the hot gas reservoir to the cold gas reservoir. 
  We assume that cold gas and stars reside in a central region of the halo with a radius of $R_s$, where $R_\mathrm{s} = R_\mathrm{vir} / c_\mathrm{dm}$ and $c_\mathrm{dm}$ is the halo's concentration. We follow the fitting functions of $c_\mathrm{dm}$ in \citet{Correa:2015aa}, which are also provided here in Appendix~\ref{appendix:cdm}. The dynamical time scale of the central region of the halo is computed with $t_\mathrm{dyn} = R_\mathrm{s} / v_\mathrm{dyn}$ and that the velocity is computed with $v_\mathrm{dyn} = \sqrt{G (M_{*}+M_\mathrm{cold}) / R_\mathrm{s}}$.\\
  
  \item {\noindent \bf Heating of cold gas and gas removal by photoionization} 
  
  We consider the effects of photoionization feedback from massive stars, $\delta M^{i}_{\rm heat}$. In haloes below the atomic cooling limit ($T_{\rm vir} = 10000$K), $\delta M^{i}_{\rm heat}$ is added directly to the outflow, whereas in haloes above the atomic cooling limit, an amount of gas with a mass of $\delta M^{i}_{\rm heat}$ is removed from the cold gas reservoir and added to the hot gas reservoir. The derivation of $\delta M^{i}_{\rm heat}$ is given in Section~\ref{sec:popiifb}. \\
  
  \item {\noindent \bf Accretion of hot gas}
  
  We assume that hot gas is continuously accreted from the intergalactic medium. In time-step $i$, an amount of hot gas
    \begin{equation}
    \begin{aligned}
        & \delta M^{i}_{{\rm acc, hot}} = \\ 
        & \left( \frac{\Omega_{\rm b}}{\Omega_{\rm m}} \mvir - M^{0}_{\rm cold} - M^{0}_{\rm hot} - M^{0}_{\rm *,II} - M^{0}_{\rm out} \right) \frac{\delta t_{i}}{\Delta t_z} , 
    \end{aligned}
    \end{equation}
    is added to the hot gas reservoir. \\

  \item {\noindent \bf Ejection of hot and cold gas}

  We account for the loss of both hot and cold gas from the halo due to the effects of stellar feedback, as described in Section~\ref{sec:popiifb}. \\

  \item {\noindent \bf Pop~II star formation}
 
  We assume that Pop~II stars form with an IMF given by \citet{Kroupa:2001aa} in the mass range $0.01-100\msun$. We sample this IMF using $4095+1$ logarithmically-spaced bins ($n_{\rm bins, II} = 4096$). Stars with masses $\leq 0.81\msun$ have stellar lifetimes larger than the age of Universe \citep{Marigo:2001aa} and they do not contribute significantly to stellar feedback. Therefore, we use only a single mass bin to represent long-lived stars and adopt a finer mass resolution for $M_{\rm star} > 0.81$ $\msun$. 
  
  To estimate how much cold gas is converted into stars during a subcycle time-step, we first calculate 
\begin{equation}
\label{eq:popiisf}
  M^{i}_{\rm *II, est} = \eta_{\rm II} M^{i}_{\rm cold} \frac{\delta t_{i}}{t^{i}_{\rm cold, ff}} , 
\end{equation}
where $\eta_{\rm II}$ is the Pop~II SF efficiency, $t^{i}_{\rm cold, ff} = (G \rho^{i}_{\rm cold})^{-1/2}$ is the free-fall time of the cold gas, $\rho^{i}_{\rm cold} = M^{i}_{\rm cold} / V_{\rm cold}$ is the mean cold gas density of the halo, $M^{i}_{\rm cold}$ and $V_{\rm cold}$ are the cold gas mass and volume that cold gas occupies, respectively. We assume that the cold gas and stars occupy only the innermost 5\% (in radius) of the halo, which is fixed during the subcycle, and so
\begin{equation*}
    V_{\rm cold} = \frac{4\pi}{3} (0.05\, R_{\rm vir})^{3}.
\end{equation*}

Next, we compute the number of stars in each IMF mass bin $j$ if they follow exactly the Kroupa IMF, $n_{j, \rm avg}$, which can be computed from $M^{i}_{\rm *II, est}$ by 
\begin{equation}
    n_{j, \rm avg} = \frac{ C_{\rm II} m_{j, \rm star}^{\alpha_{\rm II}} } { \Sigma^{N_{\rm bin}}_{j=1} C_{\rm II}m_{j, \rm{star}}^{\alpha_{\rm II}+1} } M^{i}_{\rm *II, est}, 
\end{equation}
where $m_{j, \rm{star}}$ is the stellar mass of one star in IMF bin $j$, $\alpha_{\rm II}$ is the slope and $C_{\rm II}$ is the coefficient: $C_{\rm II} = 1$ and $\alpha_{\rm II} = 0.7$ in the mass range $0.08\msun < m_{j, \rm{star}}$, $\alpha_{\rm II} = -0.3$ and $C_{\rm II}=0.08$ in the mass range $0.08\msun \leq m_{j, \rm{star}} < 0.5\msun$, and $\alpha_{\rm II} = -1.3$ and $C_{\rm II}=0.04$ in the mass range $0.5\msun \leq m_{j, \rm{star}}$ \citep{Kroupa:2001aa}. 

We calculate the averaged mass of stars with masses $< 0.81\msun$ which leads to $m_{1, \rm{star}} = 0.174\msun$ and $n_{1, \rm avg} = 2.454M^{i}_{\rm *II, est}/\msun$. We then compute 
\begin{equation}
    n_{j, \rm avg} = \frac{ 0.573(m_{j, \rm star})^{-1.3} } { \Sigma^{4096}_{j=2} (m_{j, \rm{star}})^{-0.3} } M^{i}_{\rm *II, est} 
\end{equation}
for stars with masses $\geq 0.81\msun$.

Finally, if $n_{j, \rm avg} \leq 10$, we randomly generate a number, $n_{j, \rm poi}$ of stars via Poisson sampling. Poisson sampling has the advantage that with a constant SF rate, we can still form the correct number of stars over a time-period even if the time-steps are so small that far less than one star forms per sub-time-step \citep{Sormani:2017aa}. 
We deactivate the Poisson sampling if $n_{j, \rm avg} > 10$ and use $n_{j, \rm avg}$ directly because Poisson sampling is computationally expensive and the difference between $n_{j, \rm avg}$ and $n_{j, \rm poi}$ when $n_{j, \rm avg} > 10$ is small. The total mass of newly formed Pop~II stars in each step $i$ is then $\delta M^{i}_{\rm *,II} = \Sigma m_{j, \rm star}n_{j}$, where $n_{j}$ is the number of stars in bin $j$.
Throughout the simulation, we track massive stars ($> 8\msun$) individually because their stellar feedback is important to subsequent SF. 
We adopt $\eta_{\rm II} = 2$ as our fiducial value and explain the calibration in Section~\ref{sec:cali}. 
\end{enumerate}

Finally, we adopt adaptive time-steps in the subcycles, which means that $\delta t_{i}$ varies. The adaptive time-step guarantees that all physical processes are accurately resolved when the relative change in any component of the baryonic mass budget is fast, and that we save computational time where the relative change is slow. We calculate three additional timescales:
\begin{equation}
\begin{aligned}
& t_{\rm star} = \frac{ M_{\rm *,II} }{ M_{\rm cold} } t_{\rm cold, ff}, \\
& t_{\rm cool} = \frac{M_{\rm cold}}{M_{\rm hot}} t_{\rm dyn}, \\
& t_{\rm acc} = \frac{M_{\rm hot}}{ \frac{\Omega_{\rm b}  }{\Omega_{\rm m} } \mvir - \left( M^{0}_{\rm cold} + M^{0}_{\rm hot} + M^{0}_{\rm *,II} + M^{0}_{\rm out} \right) } \Delta t_z. 
\end{aligned}
\end{equation}
Using these, we set the subcycle time-step following:
\begin{equation}
\begin{aligned}
\delta t_{i} = 0.25 \, \rm{min} (t_{\rm star}, t_{\rm cold}, t_{\rm acc}, t_{\rm dyn}, \Delta t_z).
\end{aligned}
\end{equation}
The number of subcycle time-steps ($N_\mathrm{sub}$) is then determined by the above mentioned physical processes. We make sure that $\Sigma^{N_\mathrm{sub}}_{i=1} \delta t_{i} = \Delta t_{z}$ in the code. If the computed $\delta t_\mathrm{final}$ leads to $\Sigma^{N_\mathrm{sub}}_{i=1} \delta t_{i} > \Delta t_{z}$, we then force $\delta t_\mathrm{final} = \Delta t_{z} - \Sigma^{N_\mathrm{sub}-1}_{i=1} \delta t_{i}$.
Our use of adaptive time-steps for the subcycles guarantees that mechanical and radiative feedback are sufficiently resolved in time, but significantly reduces the computational cost compared to what would be required using a small fixed time-step. 

\subsection{Stellar feedback from Pop~II stars}
\label{sec:popiifb}
During their stellar lifetime, massive stars ($> 8\msun$) emit copious amount of ionizing photons. Furthermore, stars with masses in the range $10-40\msun$ die as CCSNe \citep{Heger:2002aa,Heger:2010aa}. Each CCSN produces an amount of energy $E_{\rm SN} = 10^{51}$~erg and the energy is injected into the interstellar medium (ISM). Here we describe how the photons emitted by massive stars and the energy from SNe affect the ISM in our model. 

\begin{enumerate}
  \item {\noindent \bf{Photoheating}}
  \label{sec:photoheating}

The ionizing photons heat up the dense cold gas surrounding the star forming region. For stars with different masses, we calculate a time-averaged mass conversion rate, $\langle {\dot{M}_{\rm heat}} \rangle$, over their lifetimes. We first calculate the instantaneous mass conversion rate by 
\begin{equation}
\begin{aligned}
\dot{M}_{\rm heat} &= 10^{-25} m_{\rm H} n^{\rm den}_{\rm cold} R_{\rm D}^2 c_{\rm s} \\ 
&\left[ 1 + \frac{7}{4} \frac{c_{\rm s}(t-t_{\rm D})}{R_{\rm D}} \right]^{-1/7} ~[\msun {\rm yr}^{-1}], 
\end{aligned}
\label{eq:IoM_Main}
\end{equation}
where $m_{\rm H} = 1.66\times10^{-24}$~g is the mass of a hydrogen atom, $n^{\rm den}_{\rm cold} = 10^3$ cm$^{-3}$ is the number density, $c_{\rm s,ion} = 11.4 \left( {\rm T}_{\rm ion}/{10^4} \right)^{1/2} \times 10^5$ cm/s is the sound speed, and ${\rm T}_{\rm ion} = 10^4$K is the temperature of the ionized gas. 
We denote the distance between the ionizing front and the star, and the time it takes when the I-front reaches D-type expansion, as $R_{\rm D}$ and $t_{\rm D}$, respectively. The full derivation of Eq.~\ref{eq:IoM_Main} is in Appendix \ref{appendix:MCR}. 

We note that $R_{\rm D}^2$, $t_{\rm D}$ and therefore $\dot{M}_{\rm heat}$ show a non-linear behavior with the ionizing photon emission rate. Stars that form in a cluster are less efficient at heating their ambient medium than isolated stars. Since there is no spatial information inside the halo available from the \ctp trees, we assume that 90\% of the massive stars form in the very central region of the halo that can be considered as one big star cluster. We add the ionizing photons from these 90\% of massive stars and calculate one $\dot{M}_{\rm heat, cl}$. 
For the rest of massive stars (10\%), we assume that they form in isolation and calculate a $\dot{M}_{\rm heat, iso}$ for each of them.
In time-step $i$, we sum up the contributions from all of the massive stars to calculate the mass that is converted from cold phase to hot phase, 
\begin{equation}
    \delta M_{\rm heat, i} = \left( \dot{M}_{\rm heat, cl} + \Sigma^{N_\mathrm{iso}}_{j=1} \langle \dot{M}^{j}_{\rm heat, iso}\rangle \right) \delta t_{i}. 
\end{equation}
We examine two extreme cases: 1) all massive stars form in isolation and 2) all massive stars reside in one big star cluster and show the results in Appendix \ref{appendix:cihii}.

In mini-haloes with $T_{\rm vir} < 10000$~K, the mass heated by ionization is added directly to the outflow. To determine $T_{\rm vir}$, we use the expression
\begin{equation}
   T_{\rm vir} = \frac{G \mvir m_{\rm H}}{ R_{\rm vir} k_{\rm B} },
\end{equation}
$G = 6.67\times10^{-8} \, {\rm cm^3 \, g^{-1} \, s^{-2}}$ is the gravitational constant, $\mvir$ is the largest virial mass that the halo ever attains up to the current redshift, and $k_{\rm B} = 1.38\times10^{-16} \, {\rm cm^2 \, g \, s^{-2} \, K^{-1}}$ is the Boltzmann constant.
The reasoning behind this approximation is that these haloes are not massive enough to gravitationally bind ionized gas, as the escape velocity from these haloes is smaller than the speed of sound in the ionized gas. Sufficiently strong ionizing radiation has been shown to cause efficient outflows in such haloes, quickly removing most of the gas from them
%and quickly unbind them 
\citep{Whalen:2008, Chiaki:2018aa, Schauer:2017aa}. \citet{Visbal:2017aa} demonstrated that haloes above the atomic cooling limit can bind ionized gas and collapse under their own gravity.

  \item {\noindent \bf{Supernovae}}
  \label{sec:SNe}
  
Since we do not have spatial information on the gas inside a halo, we assume that a fraction of the gas absorbs the supernovae energy and is then unbound from the halo, i.e., gas is preferentially lost entirely rather than recycled via a galactic fountain. 
We compare the total supernovae energy deposited in the gas during time-step $i$ with the gravitational binding energy of gas in order to determine how much gas is ejected during the time-step, i.e., $\delta M^i_{\rm out, cold}$ and $\delta M^i_{\rm out, hot}$. 
The binding energy of hot gas, ${\rm E}^{i}_{\rm bind, hot}$, is a function of the dark matter mass, disk mass (i.e., cold gas mass plus stellar mass) and hot gas itself. It can be described by the following analytic equation, 
\begin{equation}
\begin{aligned}
& {\rm E}^{i}_{\rm bind, hot} = \frac{3 {\rm G} \mvir M^{i}_{\rm hot}}{R_{\rm vir} 
\left[ \frac{-R_{\rm vir}}{R_{\rm s} + R_{\rm vir}} + {\rm ln}\frac{R_{\rm s} + R_{\rm vir}}{R_{\rm s}} \right]} \times\\
& \left[ -\frac{1}{4} +\frac{1}{2} \left( 1-\frac{R_{\rm s}^2}{R_{\rm vir}^2} \right) {\rm ln}\frac{R_{\rm s}+R_{\rm vir}}{R_{\rm s}} +\frac{1}{2}\frac{R_{\rm s}}{R_{\rm vir}} \right] \\
& + \left( \frac{3R_{\rm s}}{2R_{\rm vir}}-\frac{13R_{\rm s}^3}{30R_{\rm vir}^3} \right) 
\frac{ {\rm G} M^{i}_{\rm disk} M^{i}_{\rm hot} }{R_{\rm s}} 
+ \frac{3}{5} \frac{{\rm G}(M^{i}_{\rm hot})^2}{R_{\rm vir}}, 
\end{aligned}    
\end{equation} 
where $\mvir$ is the virial halo mass a halo ever attains up to the current redshift, $R_{\rm s}$ is the scale radius of the dark matter halo, which we assume to follow an NFW profile \citep{Navarro:1996aa}, $R_{\rm vir}$ is the virial radius of the halo and $M^{i}_{\rm disk}$ is the mass of the disk. Similarly, the binding energy of cold gas can be described by 
\begin{equation}
\begin{aligned}
& {\rm E}^{i}_{\rm bind, cold} = \frac{3 {\rm G} \mvir M^{i}_{\rm cold}} {4 R_{\rm s}
\left[ \frac{-R_{\rm vir}}{R_{\rm s} + R_{\rm vir}} + {\rm ln}\frac{R_{\rm s} + R_{\rm vir}}{R_{\rm s}} \right]}  \\
& + \frac{6{\rm G} M^{\rm i}_{\rm cold} \mste}{5R_{\rm s}} + \frac{{\rm G} M^{\rm i}_{\rm hot} M^{\rm i}_{\rm cold}}{R_{\rm vir}} \left( \frac{3}{2} -\frac{3R_{\rm s}^2}{10R_{\rm vir}^2 } \right) \\
& + \frac{3{\rm G}(M^{\rm i}_{\rm cold})^2}{5R_{\rm s}}.
\end{aligned}
\end{equation}
Complete derivations of the binding energies are presented in Appendix~\ref{appendix:BE}.

We adopt an outflow efficiency with a functional form of 
\begin{equation}
\label{eq:gammaout}
    \gamma_{\rm out} =  \left( \frac{\mvir}{M_{\rm norm}} \right)^{\alpha_{\rm out}},
\end{equation}
where $M_{\rm norm}$ is the normalization mass. Both $M_{\rm norm}$ and $\alpha_{\rm out}$ are free parameters in our model. We adopt fiducial values of $M_{\rm norm} = 10^{10.5}\msun$ and $\alpha_{\rm out} = 0.72$ and explain their calibration in Section~\ref{sec:cali}. Supernovae are assumed to explode in warm, low density regions created by prior photoionization. Therefore, we first calculate how much hot gas will be removed, 
\begin{equation}
    \delta M^{i}_{\rm out, hot} = {\rm min} \left( \frac{ E^{i}_{\rm SNe}/\gamma_{\rm out} }{E^{i}_{\rm bind, hot}} M^{i}_{\rm hot}, M^{i}_{\rm hot} \right),
\end{equation} 
where $E^{i}_{\rm SNe}$ is the sum of supernovae energies that occur in this time-step.
If $E^{i}_{\rm SNe}/\gamma_{\rm out} > E^{i}_{\rm bind, hot}$, we then calculate a "leftover" supernovae energy $E^{i}_{\rm SNe, left} / \gamma_{\rm out} = E^{i}_{\rm SNe} / \gamma_{\rm out} - E^{i}_{\rm bind, hot}$. 
We only consider the ejection of cold gas if $E^{i}_{\rm SNe, left} > 0$ and obtain 
\begin{equation}
    \delta M^{i}_{\rm out, cold} = {\rm min} \left( \frac{ E^{i}_{\rm SNe, left}/\gamma_{\rm out} }{E^{i}_{\rm bind, cold}} M^{i}_{\rm cold}, M^{i}_{\rm cold} \right).
\end{equation} 
\end{enumerate}

\subsection{Ionizing and enriching volumes (internal/external enrichment)}
Reionization feedback and external enrichment are modelled according to the description in \citet{Magg:2018aa}. Each star-forming halo has an ionized bubble around it. These bubbles are launched at the virial radius and change their sizes based on the ionizing photon budget, i.e., they expand if the halo emits more ionizing photons and shrink if there are more recombination events than ionizing photons emitted. The volume $V$ of the ionized region is therefore updated from a time-step $i$ to the next time-step $i+1$ as
\begin{equation}
 V_{i+1} = \left(V_i +\frac{\dot{N}_\mathrm{ion} \Delta t}{n}\right)\left(1+\Delta t n\alpha_\mathrm{B} C\right)^{-1},
\end{equation}
where $\dot{N}_\mathrm{ion}$ is the emission rate of ionizing photons, $n$ is the mean IGM nucleon number density, $C=3$ is the clumping factor of the IGM \citep{Robertson:2013} and $\alpha_\mathrm{B} = 2.6\times10^{-13}\,\mathrm{cm}^3\,\mathrm{s}^{-1}$ is the case B recombination rate coefficient of hydrogen at $10^{4}$~K \citep{ISM_phys}. When haloes merge the volumes of the ionized bubbles are added up, as the sum conserves the number of ionizations.
 
Furthermore, each halo has a metal-enriched region around it. The expansion of the metal-enriched shell is modelled by a momentum driven snowplough, where the momentum is calculated based on the assumption that all ejected material has a constant velocity of $v_\mathrm{out} = 110\,{\rm km}\,{\rm s}^{-1}$ and slows down by sweeping up the intergalactic medium. This assumption leads to an expansion velocity of
\begin{equation}
 v_\mathrm{II} = v_\mathrm{out}\frac{M_\mathrm{out}}{M_\mathrm{out}+\frac{4}{3}\pi \rho_b\left(R_\mathrm{enr}^3-R_\mathrm{vir}^3\right)},
\end{equation}
where $\rho_b$ is the mean IGM mass density and $R_\mathrm{enr}$ is the radius of the currently enriched region around the halo. Derivations and more details of the implementation of both of the methods can be found in \citet{Magg:2018aa}.

%%%%%%%%%%%%%%%%%%%%%%%%%%%%%%%%%%%%%%%%%%%%%%%%%%%%
\section{ Calibration and parameter studies } 
\label{sec:cali}
In this section, we study the influence of the free parameters in our Pop~II SF model and describe how we determine their fiducial values. These parameters are Pop~II SF efficiency $\eta_{\rm II}$ (Eq.~\ref{eq:popiisf}), exponent $\alpha_{\rm out}$, and normalization mass $M_{\rm norm}$ in the outflow efficiency $\gamma_{\rm out}$ (Eq.~\ref{eq:gammaout}), which are defined in Section~\ref{sec:nm}.
% In addition, we consider two different slopes of our Pop~II IMF at $M_{\rm star} > 5\msun$ even though it is not a free parameter in our model, which are $\alpha_{\rm II} = -0.3$ and $\alpha_{\rm II} = -2.3$, respectively. 
In simulations other than the fiducial one, only one parameter is varied. 

\subsection{Observables used in calibration}
\label{sec:cali_obs}
\begin{enumerate}
  \item{\noindent \bf{MW properties at $z=0$}} 
  
We calibrate our model to the observed MW properties at $z=0$, which are its stellar mass $\mste$ (stars that survive until $z=0$) and cold gas mass $M_{\rm cold}$. \\
  
  \item{\noindent \bf{Cumulative stellar mass function (SMF) of the satellites}}

The SMHM relation connects the host halo and the galaxy, but it does not provide direct information on how many satellites there are in the MW system. 
Therefore, we compare the cumulative stellar mass function obtained from \asloth simulated satellites with the one from observed MW satellites \citep{McConnachie:2012aa,Munoz:2018aa}. 
To quantify the difference, we perform a two-sample Kolmogorov-Smirnov (K-S) test \citep{Kolmogorov:1933aa,Smirnov:1939aa,Massey:1951aa} on our cumulative stellar mass function and the observed one. We retrieve the KS statistic from the two-sample K-S test, which represents the maximal distance between the two stellar mass functions and is normalized to have a value between 0 and 1. The smaller the KS statistic, the more-alike the two distributions are.
Additionally, we retrieve a p-value from the same two-sample K-S test, which tells us whether we can reject the hypothesis that the two distributions are the same. The higher the p-value, the less certain we are to reject the hypothesis. \\

  \item{\noindent \bf{Stellar mass-to-halo mass relation}} 
  
We compare the SMHM relation produced by \asloth with the one derived by \citet{Garrison-Kimmel:2014ab} (hereafter GK14), who used the abundance matching (AM) to derive their SMHM relation. 
The underlying assumption of the AM technique is that the number of galaxies with stellar masses above a certain mass is the same as the number of haloes with virial masses above a certain mass \citep{Kravtsov:2004aa,Tasitsiomi:2004aa,Vale:2004aa,Conroy:2006aa,Conroy:2009aa,Guo:2010aa,Behroozi:2010aa}.
To quantify the difference between our SMHM relation and the one in GK14, we apply a mean-squared-error analysis. Due to observational completeness, the SMHM relation in GK14 is robust in the range $\mste = 10^5 - 10^8\msun$.
For \asloth simulated galaxies with stellar masses in this range, we calculate their expected stellar mass if they follow the SMHM relation in GK14, AM($\mvir$). The mean-squared-error $\chi^2$ is then 
\begin{equation}
    \chi^2 = \frac{1}{N_{\rm gal}} \sum^{N_{\rm gal}}_{i=1} \left( M_{*,i} - \rm{AM}(M_{\rm{vir, i}}) \right)^2, 
\end{equation}
where
\begin{equation}
\begin{aligned}
    &\rm{log_{10}}(\rm{AM}(\mste)) = ( \rm{log_{10}}(5.457\times10^{10}~\mste) + \\
    &f_{\rm bwc}(\rm{log_{10}}(\mste/3.266\times 10^{11}~\mste))-f_{\rm bwc}(0) ),
\end{aligned}
\end{equation}
and $M_{\rm{vir, i}}$ and $M_{*, i}$ are given in solar masses.
The fitting function $f_{\rm bwc}$ is given in \citet{Behroozi:2013aa} that 
\begin{equation}
\begin{aligned}
    f_{\rm{bwc}}(x) = &-\rm{log}_{10}(10^{\alpha x}+1) + \\
    &3.508(\rm{log}_{10}(1+\rm{exp}(x)))^{0.316/(1+\rm{exp}(10^{-x}))}.    
\end{aligned}
\end{equation}
We adopt $\alpha = 1.92$ following GK14. 
We further compare the SMHM relation produced by \asloth with the one in \citet{Nadler:2020aa} (hereafter N20), which was inferred from the fit of satellite population discovered by the Dark Energy Survey (DES) and the Panoramic Survey Telescope and Rapid Response System Pan-STARRS1 (PS1). 
\end{enumerate}

\subsection{Free parameters in the Pop~II star formation model}
\label{sec:para}
\begin{figure}
    \centering
    \includegraphics[width=\columnwidth]{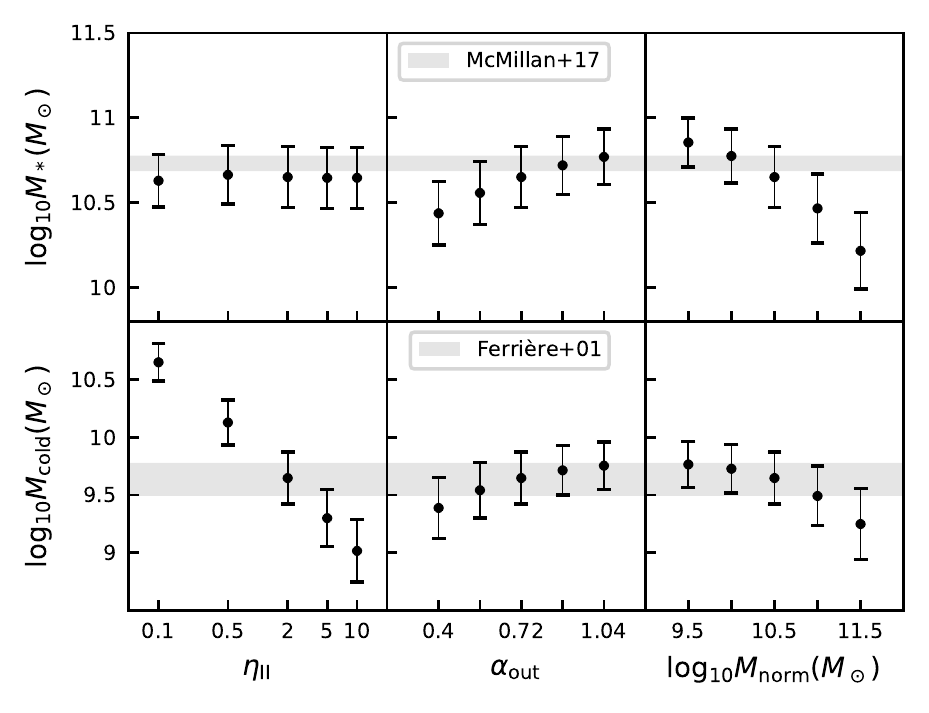}
    \caption{We show the mean values and standard deviations of $M_*$ (top panels) and $M_{\rm cold}$ (bottom panels) from 30 \ctp trees in different simulations, where we tune one of the main parameters in the Pop~II SF model. From left to right: Pop~II star formation efficiency, exponent, and normalization mass in the outflow efficiency. The upper and lower grey bands show the observational constraints: $[4.86-6] \times 10^{10}\msun$ for the observed MW stellar mass and $[3.1-6] \times 10^{9}\msun$ for the observed MW cold gas mass, respectively.}
    \label{fig:para_mstar_mcold}
\end{figure}

\begin{figure*}
    \centering
    \includegraphics{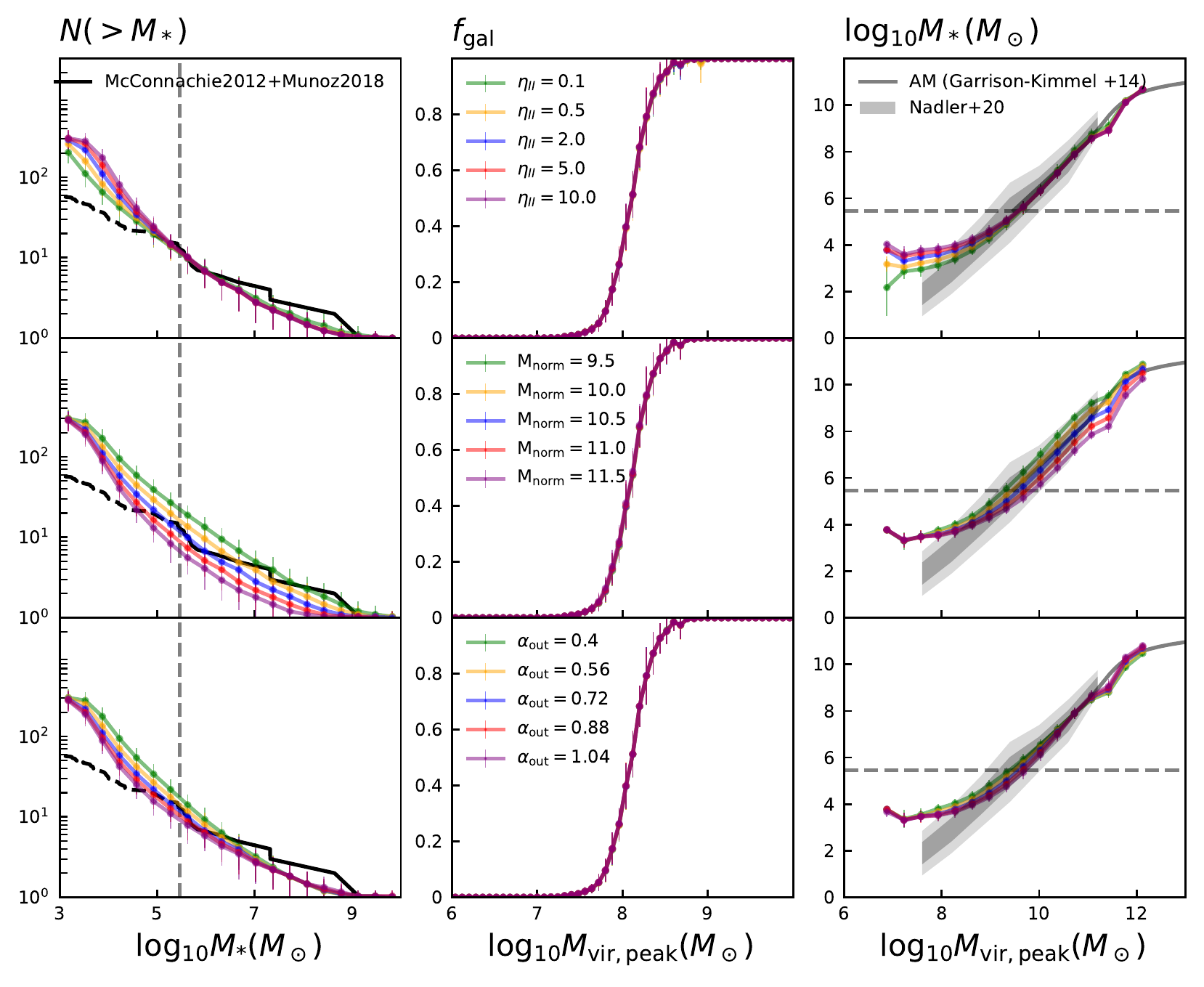}
    \caption{From left to right: cumulative SMF of satellites at $z=0$, halo occupation fraction (fraction of haloes at a given $\mvir$ that hosts a galaxy at $z=0$), and the SMHM relation at $z=0$. From top to bottom: Pop~II star formation efficiency, normalizaton mass, and exponent in the outflow efficiency. In the cumulative SMF panels, we plot the observed one \citep{McConnachie:2012aa,Munoz:2018aa} in black curve. The solid curve shows the cumulative SMF above the observational completeness, while the dashed curve shows the one below the observational completeness. In the SMHM panels, we plot the SMHM relation in GM14 in grey solid curve and the SMHM relation in N20 in grey contour. The dark grey shows $1\,\sigma$ region and the light grey contour shows the $2\,\sigma$ region of their best-fit relation. }
    \label{fig:para_smhm_cumusmf_fgal}
\end{figure*}

\begin{figure}
  \centering
  \includegraphics[width=\columnwidth]{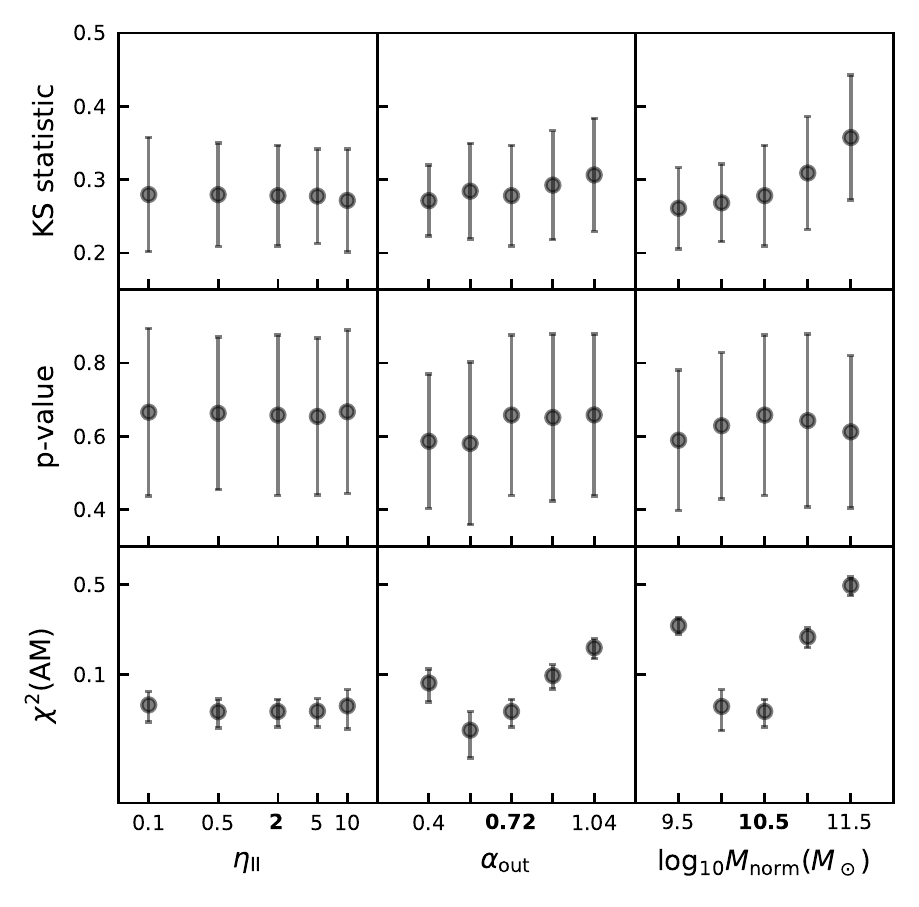}
  \caption{ Dependence of results on the main parameters. From left to right: the Pop~II SF efficiency $\eta_{\rm II}$, the exponent $\alpha_{\rm out}$ and the normalization mass $M_{\rm norm}$ in the outflow efficiency. We calibrate our model with the observed MW properties, the SMHM relation (mean-squared-error $\chi^2$), and the observed cumulative SMF (p-value and KS statistic). The mean values $\pm 1 \sigma$ (standard deviation among 30 \ctp trees) are shown and the fiducial value is highlighted with bold font. }
  \label{fig:para_cali}
\end{figure}

Here we explore how \asloth produced observables depend on the free parameters of our Pop~II SF model. These observables are the MW properties (Figure~\ref{fig:para_mstar_mcold}), the SMHM relation, halo occupation fraction (fraction of haloes at given $\mvir$ that hosts a galaxy at $z=0$), and cumulative SMF of the satellites at $z=0$ (Figure~\ref{fig:para_smhm_cumusmf_fgal}).
The quantification of differences in the observables are shown in Figure~\ref{fig:para_cali}. \\

\begin{enumerate}
  \item {\noindent \bf{Pop~II star formation efficiency}} 
  
We explore Pop~II SF efficiency $\eta_{\rm II}$ with 5 discrete values: [0.1, 0.5, 2.0, 5.0, 10.0]). 
In Figure~\ref{fig:para_mstar_mcold}, we observe that the MW stellar mass is similar among the five simulations. On the other hand, the MW cold gas mass has a clear dependence on $\eta_{\rm II}$, where the highest $\eta_{\rm II}$ gives the lowest MW cold gas mass. 
In Figure~\ref{fig:para_smhm_cumusmf_fgal}, we observe that the SMHM relations with different $\eta_{\rm II}$ are all consistent with the one in GK14 and the one in N20 above the observational completeness. Below the observational completeness, haloes with similar $\mvir$ host smaller galaxies at $z=0$ if we adopt lower $\eta_{\rm II}$.
Similarly, the difference in the cumulative SMF at $z=0$ among the five simulations only appears below the observational completeness, where the lowest $\eta_{\rm II}$ gives the fewest satellites.
In Figure~\ref{fig:para_cali}, we observe that $\eta_{\rm II}$ has a relative small influence on the p-value and $\chi^2$. This indicates that the SF in satellites is mostly feedback regulated, because they have a shallow potential. The satellites grow slower than the MW throughout the merger histories, therefore, they are less resistant to the stellar feedback. On the other hand, the MW has a much deeper potential and is more capable of retaining gas. \\

  \item{\noindent \bf{Outflow efficiency}} 

As described in Section~\ref{sec:popiifb}, the amount of gas removed by SNe is determined by comparing the binding energy of gas and the energy of SNe. We introduce $\gamma_{\rm out} = (\mvir/M_{\rm norm})^{\alpha_{\rm out}}$ such that given the same amount of SN energy, \asloth removes relatively more gas in smaller haloes than in the bigger ones. Here we examine the importance of the normalization mass $M_{\rm norm}$ and the exponent $\alpha_{\rm out}$.

We study 5 different values of $M_{\rm norm}$ spaced regularly in log-space between $10^{9.5}\msun$ and $10^{11.5}\msun$.
In contrary to $\eta_{\rm II}$, we observe that changing $M_{\rm norm}$ leads to a change in the slope of the SMHM above the observational completeness. The slope steepens when $M_{\rm norm}$ decreases, but the stellar mass in the plateau is similar among the five simulations.
We also find that adopting $M_{\rm norm} = 10^{9.5}\msun$ produces the most luminous satellite while adopting $M_{\rm norm} = 10^{11.5}\msun$ produces the least. There is a monotonic decrease in $\overline{\mste}$ and $\overline{M_{\rm cold}}$ as $M_{\rm norm}$ increases.
This is expected because higher $M_{\rm norm}$ gives lower outflow efficiency at fixed $\mvir$ and $\alpha_{\rm out}$, which means that the halo is less resistant to the SNe. 
For the exponent $\alpha_{\rm out}$, we study 5 different values distributed regularly in linear space between 0.40 and 1.04. We find that $\alpha_{\rm out}$ has an impact on the overall slope of the SMHM relation and lower $\alpha_{\rm out}$ gives a flatter SMHM relation.
When $\alpha_{\rm out} > 0$, the gas is more resistant to the SNe in haloes more massive than $M_{\rm norm}$. Thus, we observe a monotonous increase in $\overline{\mste}$ and $\overline{M_{\rm cold}}$ as $\alpha_{\rm out}$ increases.

The slope of the SMHM relation above the observational completeness is mainly influenced by $\alpha_{\rm out}$ and $M_{\rm norm}$ plays the role of the anchor point. On the other hand, the stellar mass in the SMHM plateau is mainly influenced by the Pop~II star formation efficiency $\eta_{\rm II}$. This leads to flattening of the SMHM relation occurring at different $\mvir$. 

\end{enumerate}

Based on Figures~\ref{fig:para_mstar_mcold}-\ref{fig:para_cali}, we find that the following combination of parameters gives the best overall results: $\eta_{\rm II} = 2.$, $\alpha_{\rm out} = 0.72$, and $M_{\rm norm} = 10^{10.5}\msun$. The fiducial values of our Pop~II SF model are listed in Table~\ref{table:fiducial_para}.

\begin{table}
\centering
\begin{tabular}{l|l}
     Parameter & Fiducial value \\
     \hline
     $\eta_{\rm II}$ & 2  \\
     $n_{\rm bins, II}$ & 4096 \\
     $M_{\rm bound, II}$ ($\msun$) & (0.01, 100) \\
     $M_{\rm crit}$ & S21 \\
     $\alpha_{\rm out}$ & 0.72\\
     $M_{\rm norm}$ ($\msun$) & $10^{10.5}$ \\
     $n^{\rm den}_{\rm cold}$ (cm$^{-3}$) & 1000\\
     $v_{\rm BC}$ ($\sigma_{\rm rms}$) & 0.8\\
\end{tabular}
\caption{The main parameters in our model and their fiducial values.}
\label{table:fiducial_para}
\end{table}

\section{ Results } 
In this section we discuss about the MW properties, cumulative stellar mass function, halo occupation fraction, and SMHM relation from our fiducial model in more details and the scatter among the 30 \ctp trees.\\
\label{sec:results}
\begin{figure*}
    \centering
    \includegraphics{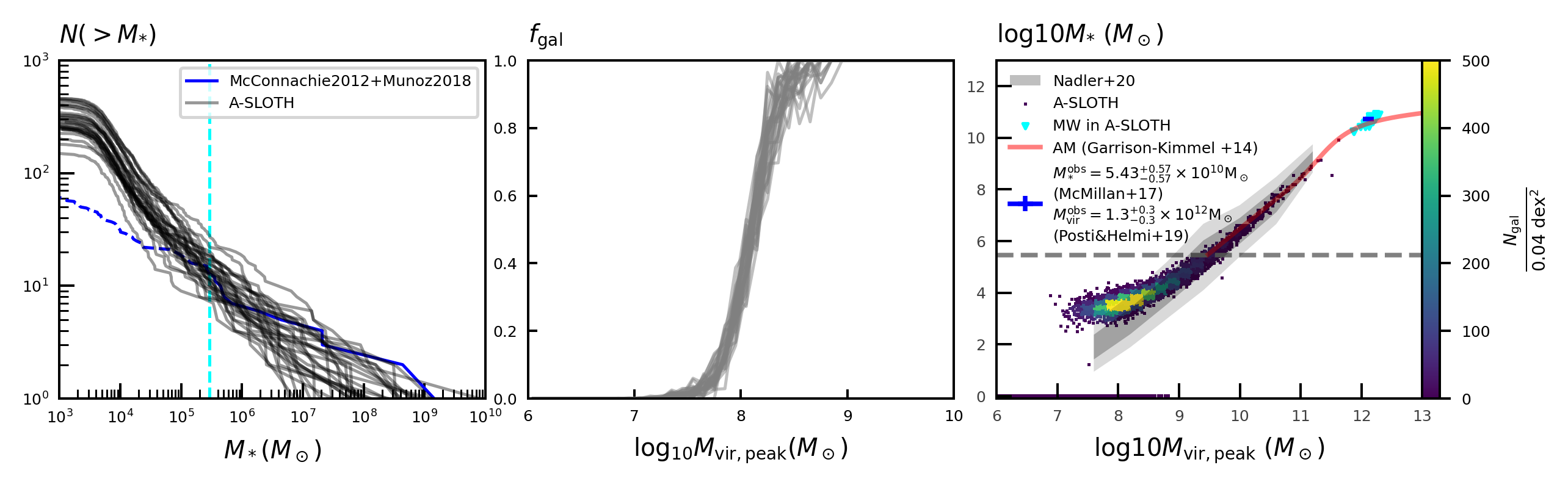}
    % [width=\columnwidth]
    \caption{
    In the left panel, \asloth simulated SMFs from 30 \ctp trees are shown in grey solid curves and the observed SMF \citep{McConnachie:2012aa,Munoz:2018aa} is plotted in blue, where the dashed curve indicates that the stellar mass is below the observational completeness (shown in the cyan dashed line). 
    In the middle panel, the halo occupation fractions from \asloth simulated galaxies are shown as grey solid curves. 
    In the right panel, each data point represents one \asloth simulated galaxy and its colour indicates the data density in the parameter space. The red solid curve represents the AM relation from GK14, which is robust at $\mste> 2.9 \times 10^5\msun$. The grey contour shows the SMHM relation from N20, where the $1\,\sigma$ contour is shown in dark grey and the $2\,\sigma$ contour is shown in light grey. The grey dashed line indicates the observational completeness and below this $\mste$, we enter the prediction region of our model. We identify the simulated MWs with cyan triangles and the blue cross marks the observed values of MW with upper and lower limits. The MW stellar mass is estimated by \citet{McMillan:2017aa} and the MW virial halo mass is estimated by \citet{Posti:2019aa}. We show satellites with zero surviving stars at the bottom of the figure.}
    \label{fig:main}
\end{figure*}

\begin{enumerate}
  \item{\noindent \bf{MW properties at $z=0$}}
  
From our fiducial model, we obtain $\overline{M_{\rm cold}} = 5\times 10^9\msun$, and $\overline{\mste} = 4.84 \times 10^{10}\msun$. The corresponding standard deviations are $\sigma_{M_{\rm cold}} = 2.29 \times10^{9}\msun$, and $\sigma_{\mste} = 1.77 \times 10^{10}\msun$, respectively. 
Our $\overline{\mste}$ is at the lower limit of the estimate by \citet{McMillan:2017aa} ($[4.86-6] \times 10^{10}\msun$). The scatter in $\mste$ among the 30 \ctp trees, $1.77 \times 10^{10}\msun$, is larger than the observational uncertainty, $0.57 \times 10^{10}\msun$. 
Since there is no distinction between atomic and molecular phase in our model, we compare our $\overline{M}_{\rm cold}$ with the combination of molecular and cold atomic masses estimated by \citet{Ferriere:2001aa}. They estimated a value in the range $[3.1-6] \times 10^{9}\msun$, which covers our result of $\overline{M_{\rm cold}} = 5 \times10^{9}\msun$. \\
  
  \item{\noindent \bf{Cumulative satellite stellar mass function}} 
  \label{sec:cumusmf} 

In Figure~\ref{fig:main}, we show the cumulative stellar mass function (SMF) of MW satellite galaxies from our fiducial model in grey and the observed one \citep{McConnachie:2012aa,Munoz:2018aa} in blue.
The scatter among the 30 \ctp trees is larger at high mass end but decreases as $\mste$ decreases. 
The number of satellites increases rapidly when stellar mass goes below the observational completeness but converges at $\approx 10^3-10^4\msun$. 
The discrepancy between the number of satellites below the observation limit is known to exist between observations and $\Lambda$ cold dark matter numerical simulations, the ``missing satellite problem'' \citep{Kauffmann:1993aa,Moore:1999aa,Klypin:1999aa}. \\

  \item {\noindent \bf{Halo occupation fraction}}
  \label{sec:fgal}
  
In the middle panel of Figure~\ref{fig:main} we plot individual halo occupation fractions (fraction of haloes at given $\mvir$ that hosts a galaxy at $z=0$) from 30 \ctp trees and find only a small scatter among them. This fraction plummets at $\mvir \approx 10^8-10^{8.5}\msun$ and stays constant while the presence and value of the stellar mass plateau changes. Thus, the presence of a plateau due to stochastically populated haloes suggested by \citet{Sawala:2015aa} is not the only reason for such a break. \\

  \item{\noindent \bf{Stellar mass-to-halo mass relation}}
  \label{sec:smhm} 

In Figure~\ref{fig:main}, \asloth simulated galaxies are plotted with coloured points, where the colour indicates the data density in a 2-dimensional phase space. The cyan triangles are the MWs in the 30 \ctp trees. 
Our fiducial model gives excellent consistency with the SMHM relation in GK14 and the one in N20.
An interesting result of our model is that we predict a flattening in the SMHM relation below $\mvir \approx 10^9\msun$. Two main factors are responsible for this plateau. 1) The outflow efficiency $\gamma_{\rm out}$, which is proportional to the peak virial mass of a halo (Eq.~\ref{eq:gammaout}). 
% The peak virial mass is defined as the maximum mass of this halo until the current redshift. 
In our model, one SN event ejects more gas in smaller haloes than in larger ones. In other words, gas is more easily retained in larger haloes, leading to more star formation. 
2) The star formation efficiency. In Figure~\ref{fig:para_smhm_cumusmf_fgal}, we observe that the turn occurs at different $\mvir$ for different $\eta_{\rm II}$ and the mass of the plateau increases when $\eta_{\rm II}$ increases. This is because the higher the star formation efficiency, the more stars form before the SNe feedback kicks in and regulate further star formation.
% In Figure~\ref{fig:SMHM_etaii_alphaout}, we show the SMHM relations from 5 simulations: $\eta_{\rm II} = 0.001$, $\eta_{\rm II} = 0.01$, Fiducial ($\eta_{\rm II} = 2$), $\eta_{\rm II} = 20$, and No $\gamma_{\rm out}$ ($\eta_{\rm II} = 2$). First, we observe that there is no obvious turn in the SMHM relation in the chase of No $\gamma_{\rm out}$. Second, we observe that the turn occurs at different $\mvir$ for different $\eta_{\rm II}$ and the mass of the plateau increases when $\eta_{\rm II}$ increases. This is because the higher the star formation efficiency, the more stars form before the SNe feedback kicks in. Therefore, the slopes at $\mvir < 10^8\msun$ are similar among the four simulations where stellar feedback and corresponding gas outflow is considered, but the turn occurs at higher $\mvir$ with higher $\eta_{\rm II}$. 
We fit the SMHM relation from our fiducial model with a broken power-law 
\begin{equation}
    \mste \propto \mvir^{1.80} \qquad \mathrm{for} \quad \mvir \geq 10^9\msun,
\end{equation}
and 
\begin{equation}
    \mste \propto \mvir^{0.66} \qquad \mathrm{for} \quad \mvir < 10^9\msun.
\end{equation}

\end{enumerate}

\subsection{ Slope of the Pop~II initial mass function }
\begin{figure*}
    \centering
    \includegraphics{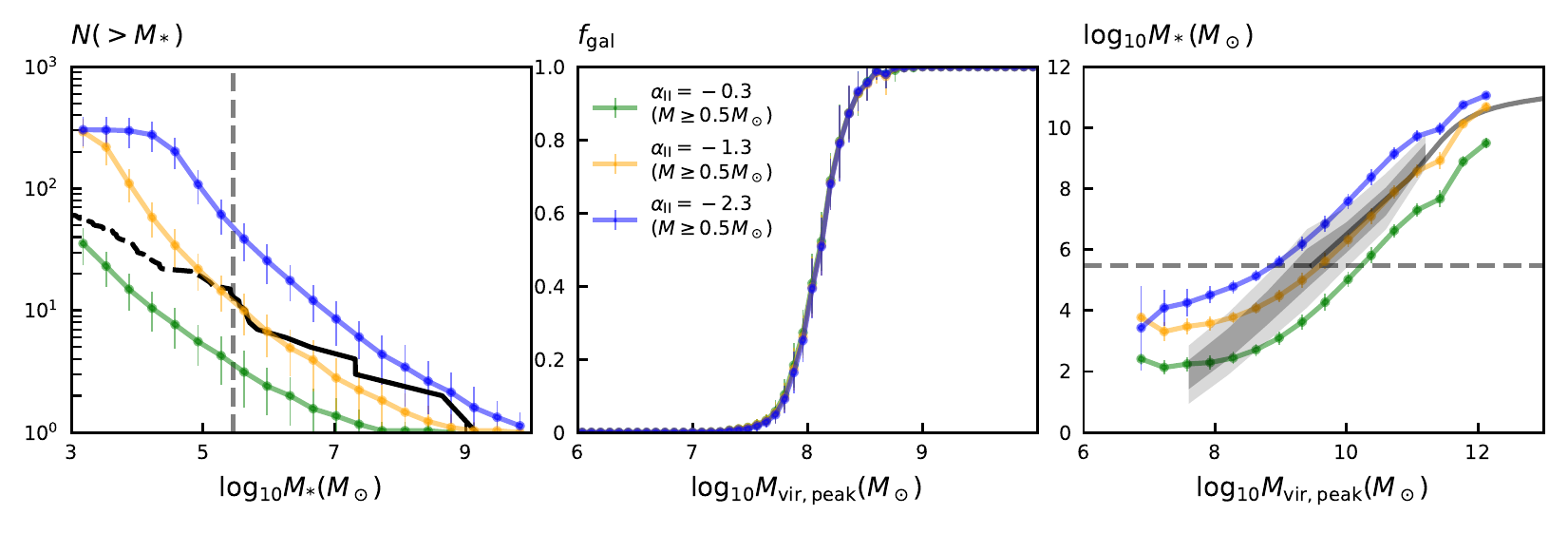}
    \caption{Similar to Figure~\ref{fig:para_smhm_cumusmf_fgal}, but we show the cumulative SMF, halo occupation fraction, and SMHM relation from different simulations where we adopt different slopes of Pop~II IMF at the high mass end ($M_{\rm star} > 0.5\msun$).}
    \label{fig:imfslope}
\end{figure*}

%  (It is important to also note that the energy feedback dominates over the stochasticity.)
Here we discuss how different Pop~II initial mass functions (IMFs) change our main results. To simplify the problem, we only change the slope of the Pop~II IMF $\alpha_{\rm II}$ in the mass range $M_{\rm star} \geq 0.5\msun$. Other than the fiducial value from the Kroupa IMF, $dN/d{\rm log}M \propto M^{-1.3}$, we examine two different slopes, which are $dN/d{\rm log}M \propto M^{-0.3}$ and $dN/d{\rm log}M \propto M^{-2.3}$, corresponding to a more top-heavy and bottom-heavy IMF, respectively. The resulting changes in the cumulative SMF, halo occupation fraction , and SMHM relation are shown in Figure~\ref{fig:imfslope}. Green, orange, and blue represent $\alpha_{\rm II} = -0.3$, $-1.3$ and $-2.3$, respectively.

We find that flattening the Pop~II IMF at $M_{\rm star} \geq 0.5\msun$ leads to lower stellar mass at $z=0$ in all haloes. The cumulative SMF is therefore below the observed one at all $\mste$. In contrary, steepening the Pop~II IMF at $M_{\rm star} \geq 0.5\msun$ makes the flattening of the cumulative SMF occur at higher $\mste$, compared to the fiducial model, and the SMF is entirely above the observed one. 
We find a universal turning point at $\mvir \approx 10^9\msun$ and that flattening the Pop~II IMF at $M_{\rm star} \geq 0.5\msun$ leads to a flatter SMHM relation at $\mvir < 10^9\msun$, but does not change the slope of SMHM relation at $\mvir \geq 10^9\msun$ significantly. 

Changing the Pop~II IMF at the high mass end is equivalent to changing the fraction of massive stars at a fixed total stellar mass. Smaller haloes ($\mvir < 10^9\msun$) are less resistant to the stellar feedback, therefore, not only the slope of the SMHM depends more heavily on the slope of the Pop~II IMF, but also the overall distribution of stellar masses of these haloes at $z=0$ depends on the slope of the Pop~II IMF, leading to different cumulative SMFs. 
% On the other hand, the slopes of the SMHM relations for larger haloes ($\mvir < 10^9\msun$) are similar among the three simulations, whereas the absolute stellar masses of these haloes still depend on the slope of the Pop~II IMF. 
The influence of different Pop~II IMFs on the halo occupation fraction is negligible.

Since we only vary the slope of the Pop~II IMF in these simulations, there is the possibility that we can tune the free parameters in our model to find another ``good'' model, which produces consistent results with the observation. However, it is not the main focus of this work to study how different Pop~II IMFs influence the results. We therefore leave this exploration for future studies.

\subsection{LW background and baryonic streaming velocity} 

\label{sec:LWstream}
As described in Section~\ref{sec:popiiisf}, the strength of the LW background ($J_{21}$) and baryonic streaming velocities ($v_{\rm BC}$, in units of root-mean-squared value $\sigma_{\rm rms}$) affect the critical mass $M_{\rm crit}$ of a halo. Here we compare different $M_{\rm crit}$ models with the fiducial one and examine the influence of $v_{\rm BC}$. 

\begin{enumerate}
    \item{ \noindent \bf{S21} } 
    
    This is our fiducial model where we follow Eqs.~9 and 10 in \citet{Schauer:2021aa} (see Section~\ref{sec:popiiisf} for details).
    The critical mass of a halo is 
\begin{equation}
    \label{eq:lw_S21}
   {\rm log_{10}} M_{\rm crit} = 6.0174 \, ( 1.0 + 0.166 \sqrt{J_{21}} ) + 0.4159 \frac{v_{\rm BC}}{\sigma_{\rm rms}},
\end{equation}
    where $J_{21}$ depends on redshift via $J_{21} = 10^{2-z/5}$. We note that we adopt constant $v_{\rm BC}$ throughout the simulation.
    
    \item {\noindent \bf{OHS}} 
    
    In our previous work, we compute $M_{\rm crit}$ by considering prescriptions in \citet{OShea:2008aa}, \citet{Hummel:2012aa}, and \citet{Stacy:2011aa}. Thus, we denote this model as OHS. 
    First, we follow \citet{OShea:2008aa} and assume that $M_{\rm crit}$ depends only on $J_{21}$. We denote this critical mass as $M_{\rm crit, O}$ that  
\begin{equation}
    M_{\rm crit, O} = 4\left( 1.25\times10^{5} + 8.7\times10^{5} \left( 4 \pi J_{21} \right)^{0.47} \right).
\end{equation} 
    Second, we follow \citet{Hummel:2012aa} and assume that when a halo reaches the critical virial temperature, $T_{\rm crit} = 2200$ K, the gas reaches high density and collapses to form Pop~III stars. We denote this critical mass as $M_{\rm crit, H}$ and compute its value  
\begin{equation}
    M_{\rm crit, H} = 10^6 \left( \frac{T_{\rm crit}}{1000 {\rm K}} \right)^{1.5} \left( \frac{1+z}{10} \right)^{-1.5}.
\end{equation}
    Finally, we follow \citet{Stacy:2011aa} and assume that the gas in a halo collapses and starts to form stars when the virial mass of the halo reaches a critical value. We denote this mass as $M_{\rm crit, S}$ and   
\begin{equation}
    M_{\rm crit, S} = \frac{\pi v_{\rm eff}^3}{6 G^{3/2} \rho^{1/2}}, 
\end{equation}
    where $v_{\rm eff} = \sqrt{v_{\rm BC}(z)^2 + c_{\rm s}^2}$ is the effective velocity of the gas, the redshift-dependent streaming velocity is given by
\begin{equation}
    v_{\rm BC}(z) = v_{\rm BC} \times \frac{6\times10^5}{201} \times (1+z),
\end{equation}
where $v_{\rm BC}$ is the input streaming velocity at $z=0$, $c_{\rm s} = \sqrt{k_{\rm B}T/\mu m_{\rm H}}$ is the sound speed, $\rho$ is the mean dark matter density of the halo and $T = 0.017(1+z)^2$K is the gas temperature \citep{Schneider:2015aa}.
    In model OHS, the final critical mass is determined by taking the maximum of $M_{\rm crit, O}$, $M_{\rm crit, H}$, and $M_{\rm crit, S}$ that 
\begin{equation}
    M_{\rm crit} = {\rm MAX}(M_{\rm crit, O}, M_{\rm crit, H}, M_{\rm crit, S}).
\end{equation}

    \item{\noindent \bf{F13}} 
    
    The last $M_{\rm crit}$ model we consider is based on the prescription in \citet{Fialkov:2013aa}. In this model F13, the virial mass of a halo needs to exceed $M_{\rm crit}$ such that the gas cools down efficiently and starts to form stars. This critical mass is computed by  
\begin{equation}
    \label{eq:f13_mcrit}
    M_{\rm crit} = M_0( 1 + 6.96(4 \pi J_{21})^{0.47} ),
\end{equation}
where $J_{21}$ depends on redshift via $J_{21} = 10^{2-z/5}$, and $M_{0}$ is the critical mass when there is no LW background. 

In \citet{Fialkov:2012aa}, they assumed that a halo starts to form Pop~III stars if its circular velocity exceeds the threshold, 
\begin{equation}
    v_{\rm cool}(z) = \sqrt{ (3.714)^2 + (4.015v_{\rm BC}(z))^2 },
\end{equation}
where the streaming velocity is dependent on redshift $v_{\rm BC}(z) = 0.0298 (1+z) v_{\rm BC}$ and $v_{\rm BC}$ is the initial streaming velocity. 
We take 
\begin{equation}
\label{eq:M_0}
    M_{0} = \left( \frac{v_{\rm cool}(z)}{146.6 \rm{km/s}} \right)^3 \times \Omega_{\rm m, 0}^{-0.5} \times (1+z)^{-1.5} \times 10^{12} {\rm h^{-1}}, 
\end{equation}
where $\Omega_{\rm m, 0} = 0.3086$ is the matter density at $z=0$. 
The full derivation of $M_0$ is in Appendix~\ref{appendix:m0}.
\end{enumerate}

Finally, as described in Section~\ref{sec:popiiisf}, we take the minimum of $M_{\rm crit}, i$ and $M_{\rm crit, 10^4{\rm K}}$ to be the final critical mass $M_{\rm crit, fin}$ in model $i$. In total we have 3 different $M_{\rm crit}$ approaches: S21, OHS, and F13. We compare the results between them while varying the initial value of streaming velocity, $v_{\rm BC} = [0, 0.8, 2, 3]\sigma_{\rm rms}$. Other parameters are fixed at the fiducial values ($\eta_{\rm II} = 2, \alpha_{\rm out} = 0.72, M_{\rm norm} = 10^{10.5}\msun$). \citet{Kulkarni:2021aa} also propose a new fitting formula for the critical mass. Due to the recency of their results, we were not able to include their proposed critical mass in our comparison.

\begin{figure*}
    \centering
    \includegraphics{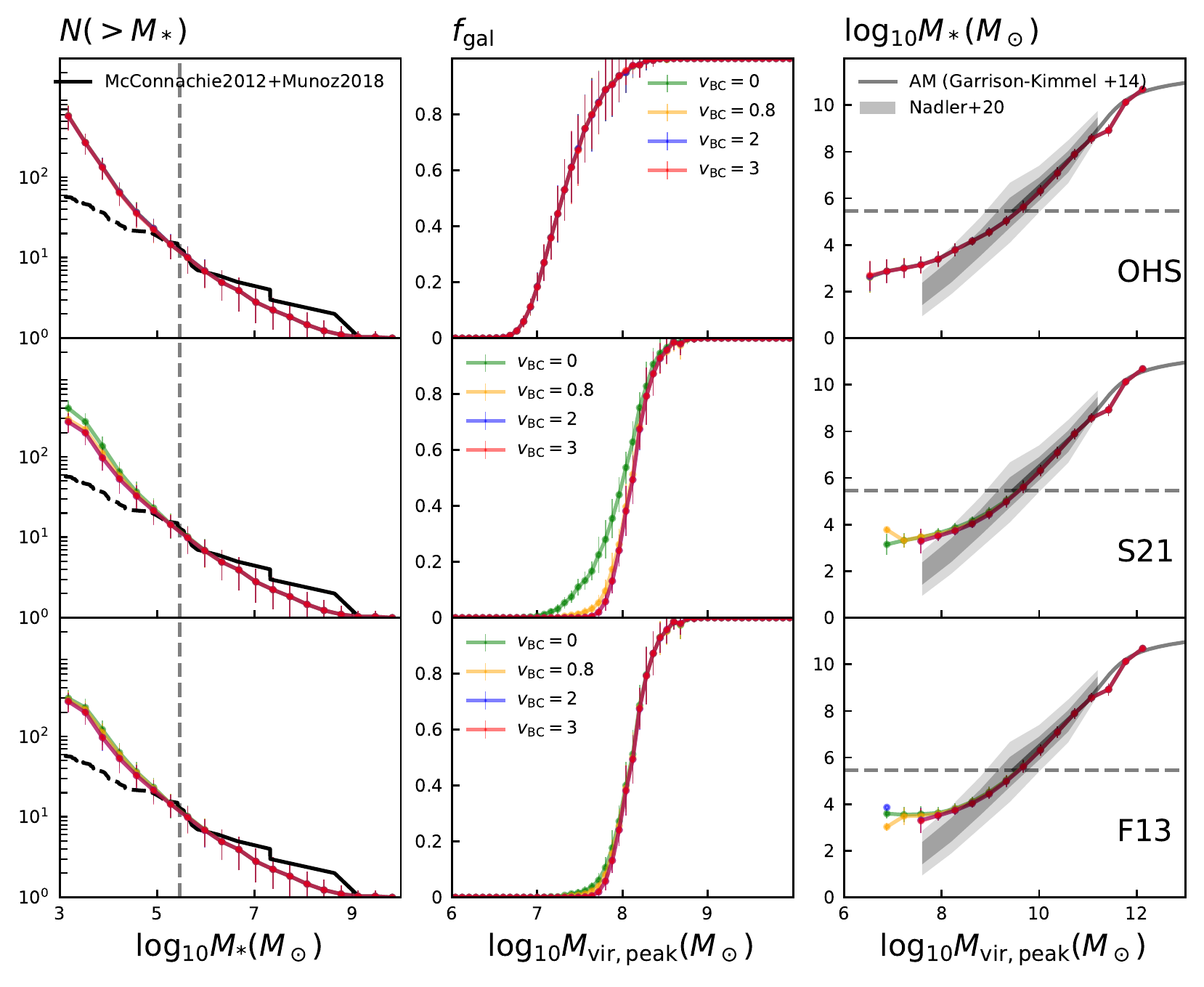}
    \caption{Similar to Figure~\ref{fig:para_smhm_cumusmf_fgal}. Here we show the cumulative SMF, halo occupation fraction, and SMHM relation from simulations where we adopt different $M_{\rm crit}$ models and different $v_{\rm BC}$. From top to bottom: OHS, S21, and F13. }
    \label{fig:lwvbc_smhm_cumusmf_fgal}
\end{figure*}

\begin{figure}
    \centering
    \includegraphics[width=\columnwidth]{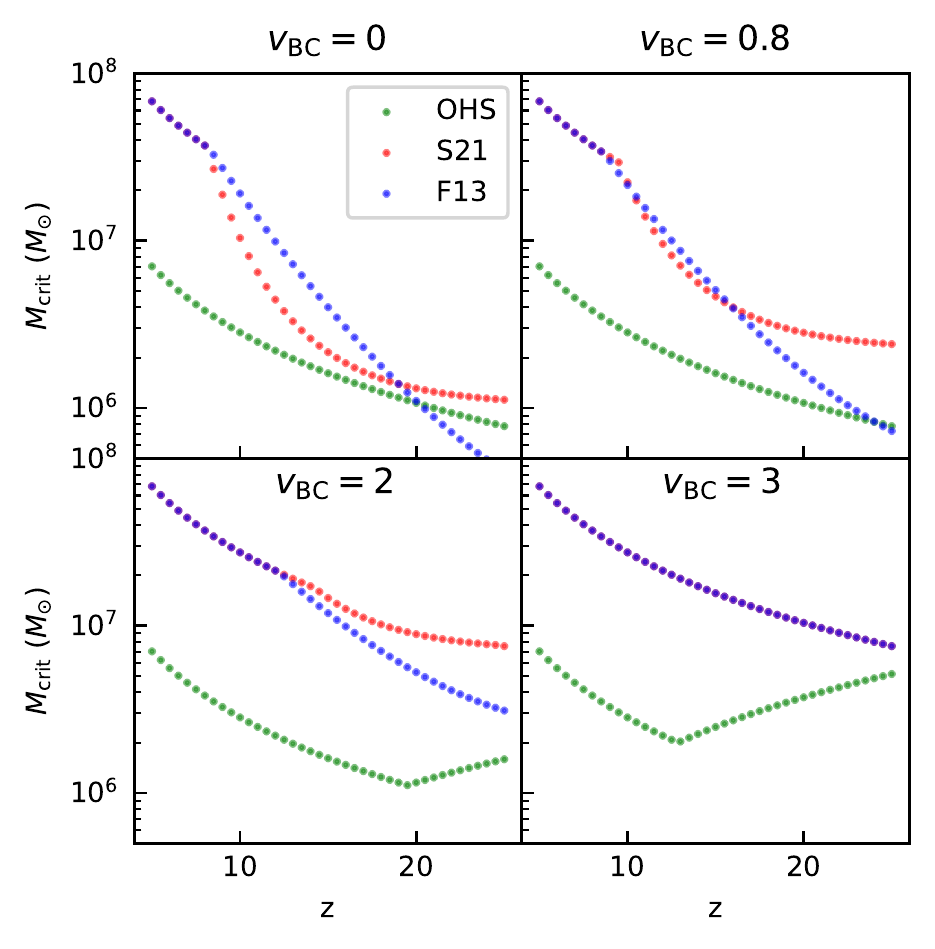}
    \caption{$M_{\rm crit}$ v.s. redshift from different $M_{\rm crit}$ models and different $v_{\rm BC}$.}
    \label{fig:mcrit}
\end{figure}

We perform the same analysis as in Section~\ref{sec:cali}.
Despite the difference in the $M_{\rm crit}$ models, the influence of LW background and streaming velocity is negligible on the mean MW stellar mass as well as the satellite-based $\chi^2$ and p-values. This indicates that we cannot tell the three $M_{\rm crit}$ models apart with the observables used in Section~\ref{sec:cali}.
We show the cumulative SMF, halo occupation fraction, and SMHM reltaion in Figure~\ref{fig:lwvbc_smhm_cumusmf_fgal}. 
The largest impact of LW background and streaming velocity is on smaller haloes, which is shown by the cumulative SMF and the halo occupation fraction. 
Model OHS produces a factor of few more ultra-faint dwarf galaxies than the other two models, regardless of the value of $v_{\rm BC}$. In model S21 we find that the number of ultra-faint dwarf galaxies $N_{\rm UFD}$ decreases as $v_{\rm BC}$ increases but saturates at $v_{\rm BC} = 2 \sigma_{\rm rms}$. 
There is a similar dependence of $N_{\rm UFD}$ on $v_{\rm BC}$ in model F13 and it produces the fewest ultra-faint dwarf galaxies among the four models when $v_{\rm BC} = 0$. 
In Figure~\ref{fig:mcrit}, we show $M_{\rm crit}$ v.s. $z$ for different $M_{\rm crit}$ models with different values of $v_{\rm BC}$. In our fiducial model ($v_{\rm BC} = 0.8 \sigma_{\rm rms}$, F13 gives the highest $M_{\rm crit}$ at $z=10-15$, followed S21, and finally OHS. This is consistent with the difference we find in $N_{\rm UFD}$.
The fraction of haloes hosting a galaxy decreases slower as $\mvir$ decreases in model OHS, whereas in models S21 and F13, this fractions experiences a plummet at $\mvir \approx 10^8\msun$.
% \tilman{Could the detection of more low-mass satellites constrain the local streaming velocity?}

%%%%%%%%%%%%%%%%%%%%%%%%%%%%%%%%%%%%%%%%%%%%%%%%%%%%%%%%
\section{Discussion}
\label{sec:dis}

\subsection{A low-mass plateau in the stellar mass-to-halo mass relation.}
\begin{figure}
    \centering
    \includegraphics[width=\columnwidth]{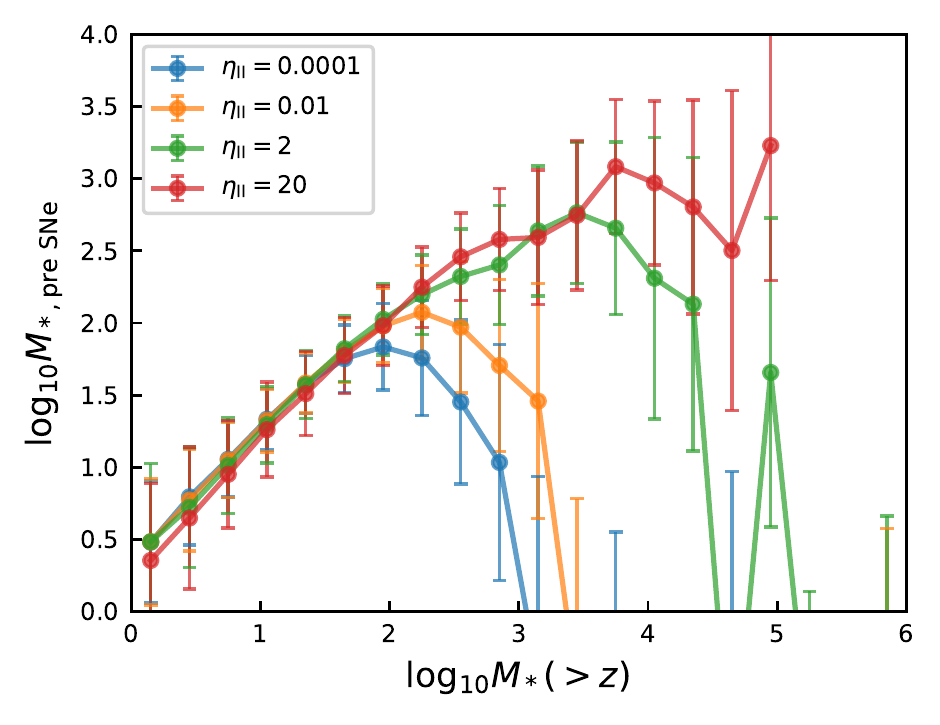}
    \caption{$M_{*, {\rm pre~SNe}}$ v.s. $M_{*}(>z)$ in haloes across all redsfhits from 1 \ctp tree. Results from different simulations with different $\eta_{\rm II}$ are plotted with different colors along with the standard deviation. Note that $M_{*, {\rm pre~SNe}}$ is computed at each redshift but $M_{*} (>z)$ is cumulative over redshifts.}
    \label{fig:mstarb4SNe}
\end{figure}

The right column of Figure~\ref{fig:para_smhm_cumusmf_fgal} shows our prediction that the SMHM does not continue as a single power-law to low masses, but instead reaches a plateau of roughly constant stellar mass.
In \asloth, this plateau is primarily due to a minimum mass scale imposed by supernova feedback: low-mass haloes able to self-quench once they form a certain amount of stars.
Figure~\ref{fig:mstarb4SNe} shows this in detail: we run simulations with different Pop~II star formation efficiencies using only 1 \ctp tree. We compute the stellar mass form before SNe occur in each global time-step ($\Delta t_z$) $M_{*, {\rm pre~SNe}}$ and the cumulative survival stellar mass $M_{*} (>z)$ for all haloes at various redshifts. 
At given survival stellar mass, we plot the mean $M_{*, {\rm pre~SNe}}$ with $1\,\sigma$ among the haloes. We observe that with higher Pop~II star formation efficiency, haloes are able to form more stars before SNe occur, which is expected. Once massive stars start to die as SNe, $M_{*, {\rm pre~SNe}}$ quickly drops.
Thus, if star formation is more efficient, more stars will form before SN feedback halts star formation.
We emphasize that this plateau is \emph{not} set by reionization in \asloth.
In this way, it may support the supernova quenching scenario proposed by earlier theoretical studies and cosmological simulations that the external ionizing background is not the dominant source to halt star formation in mini-haloes \citep{Ricotti:2005aa,Salvadori:2009aa,Bland-Hawthorn:2015aa,Jeon:2015aa,Jeon:2017aa}.
Recent observational studies also stress the importance of SN feedback based on the star formation histories of dwarf galaxies \citep{Monelli:2010aa,Monelli:2010ab,Hidalgo:2011aa,Hidalgo:2013aa,Gallart:2021aa}.
% \citet{Gallart:2021aa}, who argue that the UFD Eridanus~II has a very short star formation history consistent with a single starburst.
% [probably need some more discussion and citations here, this is not the first time it's been suggested]

A similar flattening in the SMHM relation has been observed in several previous studies, but primarily driven by reionization.
\citet{Sawala:2015aa} found such a flattening of the SMHM relation in hydrodynamic simulations, though at much higher stellar masses (${\approx}10^5\msun$). In their simulations, this effect was primarily driven by a low occupation fraction of galaxies in haloes, due to strong global reionization at $z=11$, as well as tidal stripping.
This motivated \citet{Dooley:2017aa} and \citet{Jethwa:2018aa} to consider a ``bent'' SMHM relation, which is not required to explain the MW satellite luminosity function but may alleviate tensions with the SMHM relation around the LMC (also see \citealt{Manwadkar:2021aa}).
A similar plateau is also seen at lower stellar masses (${\approx}10^3\msun$) by \citet{Wheeler:2015aa,Wheeler:2019aa}, though they emphasize this plateau could be due to resolution effects as the galaxies are composed of ${\approx}10-100$ star particles only.
More recently, \citet{Applebaum:2021aa} found such a flattening as well, but also occurring at the ${\approx}10$ star particle threshold.
Other analytic models have also seen a low-mass plateau in the SMHM relation, e.g. \citet{Kravtsov:2021aa} predict a break in the SMHM relation due to reionization and scatter in halo mass growth histories.
These previous results predict similar breaks in the SMHM as \asloth, but for a fundamentally different reason: reionization instead of supernova feedback.
Note that \asloth does include local reionization through small-scale inhomogeneities, but not yet large-scale instantaneous reionization quenching at a parameterized redshift like most of these previous models.

The presence of a stellar mass plateau has several important implications.
First, we predict that the luminosity function of UFDs should experience a sharp upturn at $\approx 10^{4}\msun$.
\citet{Dooley:2017aa} showed that such an upturn would match the luminosity function of LMC satellites, and indeed our predicted SMHM relation matches their required SMHM relation quite closely.
% (see \citealt{Manwadkar2021}).
Second, this has implications for the minimum known dark matter halo mass.
If abundance matching holds, then the lowest stellar mass UFDs should reside in the lowest mass dark matter haloes, which puts constraints on the nature of dark matter \citep[e.g.,][]{Jethwa:2018aa,Nadler:2021aa,Kim:2021aa}.
Such constraints often assume a single power-law can describe the SMHM relation at the low mass regime, but this would not be sufficient to describe the SMHM relation we predict here with \asloth.
By introducing a break in the SMHM relation, it changes the constraints obtained on dark matter properties.
Finally, we note that \asloth does not predict a very large scatter \emph{around} the low-mass end of the SMHM relation, as suggested by several previous works \citep[e.g.,][]{Garrison-Kimmel:2017aa, Munshi:2021aa}. This scatter is mainly due to star formation in some haloes being suppressed by reionization. When such large scatter is introduced, it actually \emph{steepens} the intrinsic SMHM, opposite to our prediction of a minimum stellar mass scale from SN feedback.

\subsection{Uncertainty in critical masses of haloes}

Previous studies that did not take the large-scale streaming into account 
\citep{Dijkstra:2004aa,Susa:2014aa,Visbal:2014aa,Hirano:2015aa,Skinner:2020aa} 
found that the critical mass of haloes to form stars at high redshift are typically in the range of $10^5 - 10^6 \msun$. 
% For example, \citet{Dijkstra:2004aa} utilized a one-dimensional, spherically symmetric code and found gas in haloes of masses around a few $10^6 \msun$ can cool efficiently at $z>10$ with an ionizing background. More recently, multiple authors 
In addition, \citet{ParkJ:2021aa} suggested haloes that do not form stars until $\sim 10^7 \msun$ when the LW radiation field is strong enough.
% Dijkstra2004: a few 10\^6 at z~15
% Susa2014: 10\^5-16\^6  for their star forming minihaloes Fig 17
% Hirano2015: 10\^5-16\^6 for their star forming minihaloes Sec 4.1
% Visbal2015: Figure 1, 10\^5 w/o LW feedback, 10\^6-10\^7 w/ LW feedback but no v_BC
%%% Mebane2018: Same formula as in Visbal 2014/2015
%%% Rossi2021: use the same formula as in Salvadori2009
% ParkJ2021: 10^5-10^7 with different J21 and X-ray
Studies that include large-scale streaming generally find suppression of star formation in haloes, but the minimum mass of haloes to host stars is still uncertain. 
\citet{Greif:2011aa} found the mass of star forming mini-haloes increased from $[2-3] \times 10^5 \msun$ to $~\sim 10^6 \msun$. 
\citet{Tseliakhovich:2011aa} showed that the average mass of gas-rich haloes increased from $2\times 10^4\msun$ to $2\times 10^5 \msun$ and the suppression of star formation by a factor of 1.4. 
\citet{Naoz:2013aa} concluded that the mass of haloes that are able to retain enough gas (baryonic fraction $\sim 0.1$) increased from $\sim 10^5 \msun$ to $\sim 10^6 \msun$. % with $v_\mathrm{BC} = 1\,\sigma_\mathrm{rms}$ ($5.8$~km/s at $z=199$) 
\citet{Hirano:2017ab} simulated supermassive black hole formation at high redshift and found that the virial mass of star forming haloes increased by 2 orders of magnitude at $z>30$ in their most extreme case. 
% when the initial streaming $v_\mathrm{BC}$ is set to $3\,\sigma_\mathrm{rms} (\sim 41)$~km/s at $z=499$. 
More recently, \citet{Schauer:2019aa} argued that with $v_\mathrm{BC} = 3\,\sigma_\mathrm{rms}$, star formation can be fully suppressed for haloes below the atomic cooling limit. \citet{Kulkarni:2021aa} analyse a large sample of halos from hydrodynamical simulations and find that the LW background and baryonic streaming can change the critical halos mass for Pop~III star formation by over one order of magnitude at any given redshift.
% Greif2011: Figure 3, from 2-3 X 10^5 to 6-9 X 10^5 with 3km/s at z=99
% Tseliakhovich2011: filtering mass: the minimal halo mass scale at which baryons are still able to effectively accrete on to a halo.
% Naoz2012&2013: gas fraction, didn't really claim minimum mass of a halo for star formation?
% OLeary2012: They used the function in Fialkov2012 to compute the critical mass
% Hirano2017: Table S2 shows the virial masses of their collapsing haloes, 7-26 X 10^6 with 3sigma, 1.6-55 X 10^5 w/o v_BC.
% Schauer2019: minimum mass for gas to cool, from few X 10^5 w/o v_BC to ~2 X 10^6 w/.
These studies show the importance of the relative large-scale streaming between baryon and dark matter on the star-formation process in the early Universe. There is no fully conclusive evidence on the most likely environment that MW resides in. However, we note that regions with streaming velocities of $v_\mathrm{BC} = 0.8\,\sigma$ are most likely \citep{Schauer:2021aa}, and so we adopt this in our fiducial model.

\subsection{Caveats}
The mechanical and chemical feedback of Type~Ia SNe is neglected in our current model. In the lowest-mass galaxies, Type~Ia SNe can be safely ignored because they only occur with a large delay after star formation and a couple of CCSNe is enough to quench or even suppress further star formation. In the MW, Type Ia SNe \citep{Ruiter:2009aa} are about a factor of 5 fewer than CCSNe \citep{Li:2011aa,Rozwadowska:2021aa}. Thus, we do not expect the mechanical feedback from Type Ia SNe to affect our simulations substantially. We note the importance of including Type~Ia SNe to correctly model the chemical abundances in the MW system, which we plan to improve in future studies.

We do not include an external ionization background in this model. The main source of ionization in our model is the emission of the MW. This choice is made for two reasons: First, if the modelled volume was cosmologically representative, we would find a Thomson scattering optical depth of $\tau_e \approx 0.140$ from our model. We point out that we explicitly do not try to reproduce the value from the Planck observations \citep[$\tau_e= 0.066$][]{Planck15XIII}. The fact that our optical depth is higher than the average value inferred from Planck indicates that our modelled volume experiences earlier reionization than the Universe at large and that it is dominated by local sources. Second, except for very strong ionizing radiation fields, the LW feedback completely prevents star formation in haloes below the atomic cooling limit at redshifts below $z\approx 14$. Therefore reionization feedback is only a small effect in our model.

We adopt a uniform LW background at the same redshift and ignore contributions from nearby sources. 
Earlier hydrodynamical simulations \citep{Ahn:2009aa,Johnson:2013aa} indicate that variations of the local LW intensity can span several orders of magnitudes.
Exposure to a much higher LW radiation can further suppress star formation in mini-haloes. However, this is only important for a small fraction of haloes because they need to be in close proximity to massive star-forming galaxies. 
Such spatial variation in LW radiation were considered in several semi-analytic models \citep{Agarwal:2012aa,Chon:2016aa,Visbal:2020aa} to study the formation of direct collapse black holes and Pop~III star formation rate at high redshift. 
In this work, Pop~III star formation has a minor effect on the results (number counts of MW satellites and the SMHM relation), but we plan to adopt a more detailed LW radiation model in \asloth in future studies.

\section{Conclusion}
\label{sec:sum}
We introduce an improved Pop~II SF model in the semi-analytic code \asloth, which takes dark matter merger trees as input and calculates the baryonic contents in the haloes throughout its merger history. The important physical processes include Pop~III and Pop~II SF, aw well as mechanical and chemical feedback from massive stars. The new feature in the Pop~II SF model is that we are now able to trace individual Pop~II stars, which allows us to precisely resolve stellar feedback in space and time. We take 30 merger trees from the \ctp project \citep{Griffen:2016aa} and use \asloth to study the SMHM relation at $z=0$ of satellites in MW-like systems. 

In this work, we focus on the Pop~II SF model. The main free parameters are the star formation efficiency $\eta_{\rm II}$, normalization mass $M_{\rm norm}$ and exponent $\alpha_{\rm out}$ in the outflow efficiency $\gamma_{\rm out}$. We calibrate our model with the observed stellar mass \citep{McMillan:2017aa} and cold gas mass of MW 
\citep{Ferriere:2001aa}, the SMHM relation derived with AM technique from GK14 and N20 above the observational completeness, and the observed cumulative stellar mass function of satellite galaxies \cite{McConnachie:2012aa,Munoz:2018aa}. 
We find $\eta_{\rm II} = 2$, $M_{\rm norm} = 10^{10.5} \msun$, and $\alpha_{\rm out} = 0.72$ gives the most consistent results with the observation and adopt these as fiducial values. 

Our fiducial model produces a mean MW stellar mass of 4.84$\times 10^{10}\msun$ and mean cold gas mass of 5$\times 10^{9}\msun$, which are consistent with the observed values of $[4.86-6] \times 10^{10}\msun$ and $[3.1-6] \times 10^9\msun$, respectively. 
The cumulative stellar mass function of MW satellite galaxies from \asloth simulation is consistent with the observation above the observational completeness. We also find that the cumulative SMF has an upturn when $\mste$ is below observational completeness and the difference between merger trees decreases. Below $\mste\approx 10^3-10^4\msun$ the cumulative SMF flattens. 
The fraction of haloes that host a galaxy (halo occupation fraction) increases rapidly above $\mvir \approx 10^{8}\msun$, and it is insensitive to the free parameters in our Pop~II SF model. 
Our fiducial SMHM relation is consistent with the ones in GK14 and N20. We find that the slope of the SMHM relation for stellar masses is determined by the exponent $\alpha_{\rm out}$ and normalization mass $M_{\rm norm}$ in the outflow efficiency. We also observe a plateau in the SMHM relation at $\mvir \leq 10^{9}\msun$. This plateau represents a minimum stellar mass that forms before SNe occur and quench further star formation, which is mainly determined by the Pop~II star formation efficiency $\eta_{\rm II}$. 

We further examine how $\alpha_{\rm II}$, the slope of Pop~II IMF at the high mass end ($M_{\rm star} > 0.5\msun$), influences the results. For a top-heavy IMF, with $\alpha_{\rm II}$ below the \citet{Kroupa:2001aa} value, the fraction of SNe at given stellar mass is lower, leading to weaker feedback. On the contrary, for a more bottom-heavy IMF, the fraction of SNe at given stellar mass is higher and the resulting stellar feedback is stronger. Therefore, galaxies have higher stellar mass in simulations with steeper $\alpha_{\rm II}$ than in those with shallower $\alpha_{\rm II}$.
Finally, we examine the impact of different $M_{\rm crit}$ models and of the streaming velocity. We compare three different $M_{\rm crit}$ approaches: 1) the default model from Eqs.~9 and 10 in \citet{Schauer:2021aa}, which we denote as S21, 2) a combination of $M_{\rm crit}$ formulae from \citet{OShea:2008aa}, \citet{Hummel:2012aa}, and \citet{Stacy:2011aa}, which we denote as OHS, and 3) $M_{\rm crit}$ formula from \citet{Fialkov:2013aa}, which we denote as F13. We find that model OHS produces a factor of few more ultra-faint dwarf galaxies and the effect of streaming velocity is negligible. Models S21 and F13 show similar dependence on streaming velocity that $N_{\rm UFD}$ decreases as $v_{\rm BC}$ increases and stays constant after $v_{\rm BC} = 2 \sigma_{\rm rms}$. In model OHS, the halo occupation fraction experiences a slower decrease as $\mvir$ decreases, whereas in models S21 and F13, the halo occupation fraction expreiences a plummet at $\mvir \approx 10^8\msun$.

The newly implemented Pop~II SF model in \asloth provides us an efficient and reliable framework to follow the formation of individual stars and their corresponding feedback at appropriate timings. This new model also enables us to examine the properties of surviving stars individually. We plan to perform follow-up studies on the properties of metal-poor stars in MW dwarf satellite galaxies at $z=0$. 

\section*{Acknowledgements}
We thank Yuta Tarumi and Anna T. P. Schauer for useful discussion and Ethan O. Nadler for providing data points to reproduce their SMHM relation. We also thank the referee for carefully reading the manuscript and providing valuable feedback. We gratefully acknowledge the HPC resources and data storage service SDS@hd supported by the Ministry of Science, Research and the Arts Baden-Württemberg (MWK) and the German Research Foundation (DFG) through grant INST 35/1314-1 FUGG and INST 35/1503-1 FUGG. LHC, SCOG, and RSK acknowledge financial support from DFG via the Collaborative Research Center (SFB 881, Project-ID 138713538) 'The Milky Way System' (subprojects A1, B1, B2, B8). 
%SCOG and RSK acknowledge support from the DFG via the collaborative research centre (SFB 881, Project-ID 138713538) “The Milky Way System” (sub-projects A1, B1, B2 and B8) 
SCOG and RSK also thank for funding from the Heidelberg cluster of excellence (EXC 2181 - 390900948) `STRUCTURES: A unifying approach to emergent phenomena in the physical world, mathematics, and complex data', and from the European Research Council in the ERC synergy grant `ECOGAL – Understanding our Galactic ecosystem: From the disk of the Milky Way to the formation sites of stars and planets' (project ID 855130). TH acknowledges funding from JSPS KAKENHI Grant Numbers 19K23437 and 20K14464.

\section*{DATA AVAILABILITY}
The data underlying this article will be shared on reasonable
request to the corresponding author.

%%%%%%%%%%%%%%%%%%%%%%%%%%%%%%%%%%%%%%%%%%%%%%%%%%



\bibliographystyle{mnras}
\bibliography{Reference.bib}

%%%%%%%%%%%%%%%%%%%%%%%%%%%%%%%%%%%%%%%%%%%%%%%%%%

%%%%%%%%%%%%%%%%% APPENDICES %%%%%%%%%%%%%%%%%%%%%

\appendix

%Some extra material
% If you want to present additional material which would interrupt the flow of the main paper,
% it can be placed in an Appendix which appears after the list of references.
\section{Choice of number of Pop~II IMF bins}
\label{appendix:nbinII}
We show the averaged number of ionizing stars and SNe we get out of every 1000$\msun$ of stars in our model in Table\ref{table:nbins} with different number of Pop~II IMF bins $n_{\rm bins, II}$. The difference between 4096 and 8192 bins is less than 0.1\% in N$_{\rm ion}$ and identical in N$_{\rm SNe}$. Therefore, we adopt $n_{\rm bins, II} = 4096$ in our model.
\begin{table}
    \centering
    \begin{tabular}{c|c|c}
        $n_{\rm bins, II}$ & N$_{\rm ion}$ & N$_{\rm SNe}$ \\
        128 & 10.69 & 7.85 \\
        256 & 10.55 & 7.75 \\
        512 & 10.48 & 7.71 \\
        1024 & 10.51 & 7.73 \\
        2048 & 10.49 & 7.74 \\
        4096 & 10.49 & 7.74 \\
        8192 & 10.48 & 7.74 \\ 
        16384 & 10.48 & 7.74
    \end{tabular}
    \caption{Averaged number of ionizing stars and SNe that we draw from the IMF given a total stellar mass of 1000$\msun$.}
    \label{table:nbins}
\end{table}

\section{Detailed derivation of mass conversion rate and gas binding energy}
\subsection{Mass conversion rate}
\label{appendix:MCR}
Here we describe the derivation of Eq.~\ref{eq:IoM_Main}. We first calculate the mass which is enclosed in the HII region of a D-type I-front, which can be described as a function of $t$ 
\begin{equation}
\begin{aligned}
M_{\rm HII} = m_{\rm H} \frac{4 \pi}{3} R_{\rm I}^3(t) n_{\rm HII}(t) ~({\rm g}), \\
\end{aligned}
\end{equation}
where $n_{\rm HII}(t)$ is the number density at time $t$ inside the HII region and $m_{\rm H}$ is the atomic mass. For simplicity, we ignore helium.
The number density inside the HII region is computed by 
\begin{equation}
  n_{\rm HII}(t) = n^{\rm den}_{\rm cold} \left( \frac{R_{\rm D}}{R_{\rm I}(t)} \right)^{3/2} \rm{(cm^{-3})}, 
\end{equation}
where $n^{\rm den}_{\rm cold}$ is the number density of the dense gas. The radius that the I-front reaches at time $t$, $R_{\rm I}(t)$ is computed by \citep{Spitzer:1978aa}
% \begin{equation}
% \begin{aligned}
% n_{\rm HII}(t) = \left( \frac{Q}{4 \pi \alpha_{\rm B} R_{i}(t)} \right)^{1/2}~[{\rm cm}^{-3}],
% \end{aligned}
% \end{equation}
% where $Q$ is the rate at which the central source produces ionizing photons, measured in units of photons per second. 
\begin{equation}
\begin{aligned}
R_{\rm I}(t) = R_{\rm D} \left[ 1 + \frac{7}{4} \frac{c_{\rm s}(t-t_{\rm D})}{R_{\rm D}} \right]^{4/7} (\rm{cm}), 
\end{aligned}
\end{equation}
where $c_{\rm s} = 11.4$ $\left( \frac{{\rm T}_{\rm ion}}{T_4} \right)^{1/2} \times 10^5$cms$^{-1}$ is the sound speed and we assume for simplicity a fixed temperature $T_{\rm ion} = 10^4$K. 
The distance that the I-front reaches when it enters D-type expansion, $R_{\rm D}$, is 
\begin{equation}
\begin{aligned}
R_{\rm D}= \left( \frac{3 Q}{4 \pi (n^\mathrm{den}_\mathrm{cold}})^2 \alpha_{\rm B} \right)^{1/3} \rm (cm),
\end{aligned}
\end{equation}
and the time it needs to reach this distance, $t_{\rm D}$, is given by 
\begin{equation}
\begin{aligned}
t_{\rm D} = {\rm ln}\left(\frac{842}{23} \left(\frac{Q}{10^{48}}\right)^{1/3} \left(\frac{n^{\rm den}_{\rm cold}}{10^3}\right)^{1/3}\right) \tau (\rm{s}), 
\end{aligned}
\end{equation}
where $\tau = (\alpha_{\rm B}n^{\rm den}_{\rm cold})^{-1}$s is the recombination time. 
% We compute the time derivative of $R_{\rm{i}}(t)$, 
% \begin{equation}
% \begin{aligned}
% \frac{d R_{\rm i}(t)}{dt} = c_{\rm s} \left[ 1 + \frac{7}{4} \frac{c_{\rm s}(t-t_{\rm D})}{R_{\rm D}} \right]^{-3/7} \rm{(cm/s)}.
% \end{aligned}
% \end{equation} 
We can then express $M_{\rm HII}(t)$ as
\begin{equation}
\begin{aligned}
&M_{\rm HII}(t) \\
&=  \frac{4 \pi m_{\rm H}}{3} R_{\rm D}^{3} \left[ 1 + \frac{7}{4} \frac{c_{\rm s}(t-t_{\rm D})}{R_{\rm D}} \right]^{12/7}  n^{\rm den}_{\rm cold} \left( \frac{R_{\rm D}}{R_{\rm I}(t)} \right)^{3/2} \\
&=m_{\rm H} n^{\rm den}_{\rm cold} \frac{4 \pi}{3} R_{\rm D}^{3} \left[ 1 + \frac{7}{4} \frac{c_{\rm s}(t-t_{\rm D})}{R_{\rm D}} \right]^{6/7} ({\rm g}).
\end{aligned}
\end{equation}

\begin{figure}
  \centering
  \includegraphics[width=\columnwidth]{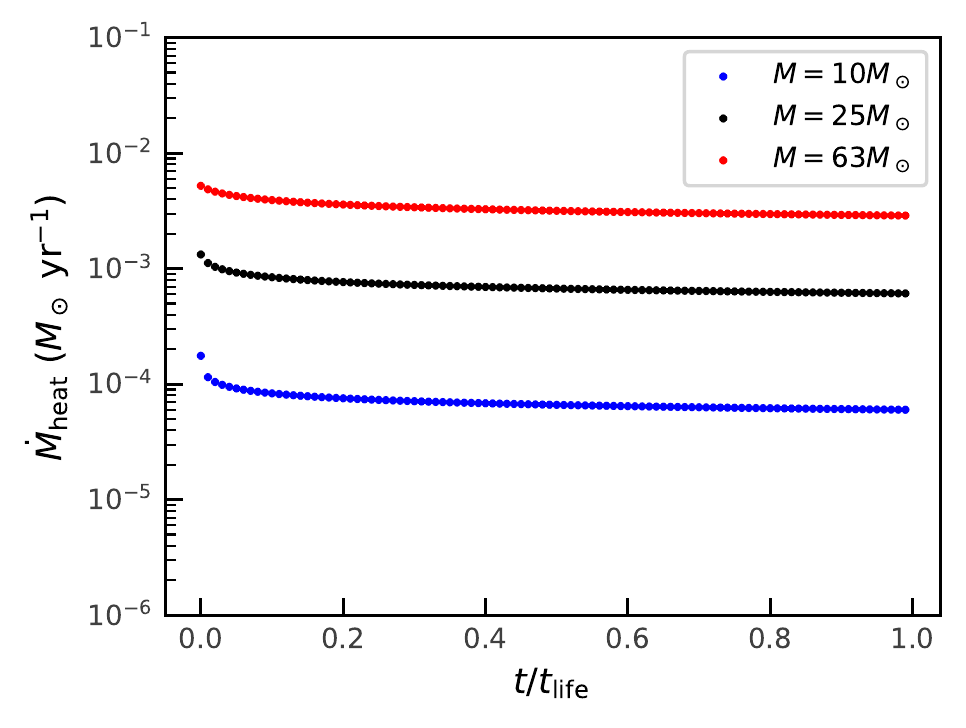}
  \caption{Mass conversion rate of 10, 25, and 63 $\msun$ stars throughout their lifetimes. These values are computed using Equation~\ref{mdotconv} with an assumed cold gas density of $n^{\rm den}_{\rm cold} =  1000 \, {\rm cm}^{-3}$.}
  \label{fig:IoM}
\end{figure}

Finally, we can use the expression above to write down the rate at which gas is incorporated into the HII region, converting it from the cold phase in our model to the hot phase. This mass conversion rate is given as a function of time, 
$Q$ and $n^{\rm den}_{\rm cold}$ by
\begin{equation}
\begin{aligned}
\dot{M}_{\rm heat} &= 10^{-25} m_{\rm H} R_{\rm D}^2 c_{\rm s} \left[ 1 + \frac{7}{4} \frac{c_{\rm s}(t-t_{\rm D})}{R_{\rm D}} \right]^{-1/7} \\ & n^{\rm den}_{\rm cold} (\msun {\rm yr}^{-1}).
\label{mdotconv}
\end{aligned}
\end{equation}
As an example, we show the mass conversion rates for 10, 25 and 63$\msun$ stars over the course of their lives in Figure~\ref{fig:IoM}. These values were computed assuming a cold gas density $n^{\rm den}_{\rm cold} = 1000$~cm$^{-3}$, but are only weakly sensitive to this choice. 

After a decrease in a small fraction of the stellar lifetime, $\dot{M}_{\rm heat}$ stays almost constant until the end of the stellar lifetime. 
Based on Table~6 in \citet{Schaerer:2002aa}, we find that the ionizing photon rate $Q$ of massive stars (7-150$\msun$) can be well described with the following equation, 
\begin{equation}
\begin{aligned}
{\rm log}_{10}Q = &~27.8 + 30.68 \left(\frac{M_{\rm star}}{\msun}\right) \\
&- 14.8 \left(\frac{M_{\rm star}}{ \msun }\right)^2 + 2.59 \left(\frac{M_{\rm star}}{\msun}\right)^3.
\end{aligned}
\end{equation}

\subsection{Fitting function of halo concentration}
\label{appendix:cdm}
We compute the scale radius of a halo $R_\mathrm{s}$ in (Section~\ref{sec:popiifb}) following the fitting equations of halo concentration in \citet{Correa:2015aa}:\\
when $z < 4$,
\begin{equation}
\begin{aligned}
    &\mathrm{log}_{10} c_\mathrm{dm} = a + b \mathrm{log}_{10}(M_\mathrm{vir})\left(1+cM_\mathrm{vir}^2\right) \\
    &a = 1.7543 - 0.2766(1+z) + 0.02039(1+z)^2 \\
    &b = 0.2753 + 0.00351(1+z) - 0.3038(1+z)^{0.0269} \\
    &c = -0.01537 + 0.02102(1+z)^{-0.1475} \\
\end{aligned}
\end{equation}
and when $z\geq4$,
\begin{equation}
\begin{aligned}
    &\mathrm{log}_{10} c_\mathrm{dm} = a + b \mathrm{log}_{10}(M_\mathrm{vir}) \\
    &a = 1.3081 - 0.1078(1+z) + 0.00398(1+z)^2 \\
    &b = 0.0223 - 0.0944(1+z)^{-0.3907}.
\end{aligned}
\end{equation}

\subsection{Gas binding energy}
Here we show the full derivation of cold and hot gas binding energies that are used to determine the outflow mass in Section~\ref{sec:nm}. Without spatial information of the baryons, we assume a uniform density inside each baryonic content. There are four components that contribute to the total binding energy: the dark matter halo, the cold gas, the hot gas and the stars. Stars and cold gas reside in the same region, that we define as the disk (innermost 5$\%$ region in terms of radius). The contribution from each component can be calculated separately and summed up later to give Eqs.~\ref{eq:BEHotTotal} and \ref{eq:BEColdTotal}.
\label{appendix:BE}

\begin{equation}
\label{eq:BEHotTotal}
\begin{aligned}
  & E_{\rm bind, hot} = \int_{0}^{R_{\rm vir}} dM_{\rm hot} U(r) \\
  & = E_{\rm hot,DM} + E_{\rm hot,disk} + E_{\rm hot,hot} \\
  & = \int_{0}^{R_{\rm vir}} dM_{\rm hot}  U_{\rm DM}(r) + \int_{0}^{R_{\rm vir}} dM_{\rm hot}  U_{\rm disk}(r) \\ 
  & + \int_{0}^{R_{\rm vir}} dM_{\rm hot} U_{\rm hot}(r),
\end{aligned}
\end{equation}
where $U$ is the potential energy.

\begin{equation}
\label{eq:BEHotHot}
\begin{aligned}
& {\rm E}_{\rm hot, hot} = \int_0^{R_{\rm vir}} \frac{{\rm G} M_{\rm hot}(r) dM_{\rm hot}}{R}  \\
& = \int_0^{R_{\rm vir}} \frac{{\rm G}\frac{4}{3} \pi r^3 \rho_{\rm hot} 4 \pi r^2 \rho_{\rm hot} dr }{r} \\
% & = \int_0^{R_{\rm vir}} {\rm G}\frac{16 \pi^2 r^4 \rho_{\rm hot}^2}{3} dr \\
% & = \frac{16}{15} \pi^2 {\rm G} \rho_{\rm hot}^2 r^5 \Bigg\rvert^{R_{\rm vir}}_0 \\
% & = \frac{16}{15} \pi^2 {\rm G} \rho_{\rm hot}^2 R_{\rm vir}^5 
& = \frac{3}{5} \frac{{\rm G}M_{\rm hot}^2}{R_{\rm vir}}. \\ 
\end{aligned}
\end{equation}

\begin{equation}
\label{eq:BEHotDisk}
\begin{aligned}
&{\rm E}_{\rm hot, disk} = \int_0^{R_{\rm vir}} dM_{\rm hot} \int_r^{\infty} \frac{{\rm G}M_{\rm disk}(R)}{{R}^2} dR \\
& = \int_0^{R_{\rm s}} dM_{\rm hot} \left[ \int_r^{R_{\rm s}} \frac{{\rm G}M_{\rm disk}(R)}{{R}^2} dR +\int_{R_{\rm s}}^\infty \frac{{\rm G}M_{\rm disk}}{{R}^2} dR \right] \\
& + \int_{R_{\rm s}}^{R_{\rm vir}} dM_{\rm hot} \frac{{\rm G}M_{\rm disk}}{{r}} \\
% & = 4 \pi \rho_{\rm hot} \left[ {\rm G} \frac{2 \pi (\dfrac{r^3}{3}R_{\rm s}^2 - \dfrac{r^5}{5}) \rho_{\rm disk}}{3} + \frac{r^3}{3}\frac{{\rm G}M_{\rm disk}}{{R_{\rm s}}} \right] \Bigg\rvert^{R_{\rm s}} _0\\
% & + \int_{R_{\rm s}}^{R_{\rm vir}} dM_{\rm hot} \frac{{\rm G}M_{\rm disk}}{{r}} \\
& = 4 \pi \rho_{\rm hot} \left[ {\rm G} \frac{2 \pi (\frac{R_{\rm s}^3}{3}R_{\rm s}^2 - \frac{R_{\rm s}^5}{5}) \rho_{\rm disk}}{3} + \frac{R_{\rm s}^3}{3}\frac{{\rm G}M_{\rm disk}}{{R_{\rm s}}} \right] \\
& + 2 \pi \rho_{\rm hot} (R_{\rm vir}^2 - R_{\rm s}^2) G M_{\rm disk} \\
% & = \frac{R^3_{\rm s}}{15R^3_{\rm vir}} \frac{G M_{\rm disk}M_{\rm hot}}{R_{\rm s}} + \frac{R^3_{\rm s}}{R^3_{\rm vir}} \frac{G M_{\rm disk}M_{\rm hot}}{R_{\rm s}} \\
% & + M_{\rm hot} \frac{3}{2} ( \frac{R_{\rm s}}{R_{\rm vir}} - \frac{R_{\rm s}^3}{R^3_{\rm vir}} ) G M_{\rm disk} \\
& = \left(\frac{3R_{\rm s}}{2R_{\rm vir}}-\frac{13R_{\rm s}^3}{30R_{\rm vir}^3}\right) \frac{{\rm G}M_{\rm hot} M_{\rm disk}}{{R_{\rm s}}}, \\
\end{aligned}
\end{equation}
where $M_{\rm disk} = M_{\rm stellar} + M_{\rm cold}$.

\begin{equation}
\label{eq:BEHotDM}
\begin{aligned}
&{\rm E}_{\rm hot, DM} = \int_0^{R_{\rm vir}} dM_{\rm hot} \int_r^{\infty} \frac{{\rm G}M_{\rm DM}(R)}{{R}^2} dR \\
% & = \int_0^{R_{\rm vir}} dM_{\rm hot} \int_r^{\infty} \frac{{\rm G}}{{R}^2} dR \int_0^{R} 4 \pi {r'}^2 \rho(r') dr' \\
&= \int_0^{R_{\rm vir}} dM_{\rm hot} \int_r^{\infty} \frac{{\rm G}}{{R}^2} dR \int_0^{R} 4 \pi {r'}^2 \frac{\rho_0}{\dfrac{r'}{R_{\rm s}} \left(1 + \dfrac{r'}{R_{\rm s}} \right)^2} dr' \\
& = 4 \pi {\rm G} \rho_0 R_{\rm s}^3 \int_0^{R_{\rm vir}} dM_{\rm hot}  \int_r^{\infty} \frac{dR}{{R}^2} \int_0^{R} \frac{{r'}}{ \left(R_{\rm s} + r' \right)^2} dr' \\
% & = 4 \pi {\rm G} \rho_0 R_{\rm s}^3 \int_0^{R_{\rm vir}} dM_{\rm hot}  \int_r^{\infty} \frac{dR}{{R}^2} \left[ \frac{R_{\rm s}}{R_{\rm s}+r'} + {\rm ln}\left(R_{\rm s}+r'\right) \right] \Bigg\rvert^R_0 \\
% & = 4 \pi {\rm G} \rho_0 R_{\rm s}^3 \int_0^{R_{\rm vir}} dM_{\rm hot} \int_r^{\infty} \frac{ \left[ \frac{R_{\rm s}}{R_{\rm s} + R} + {\rm ln}\frac{\left( R_{\rm s} + R\right)}{R_{\rm s}} - 1 \right] }{R^2} dR \\
% & = 4 \pi {\rm G} \rho_0 R_{\rm s}^3 \int_0^{R_{\rm vir}} 4 \pi r^2 \rho_{\rm hot} \left( \frac{{\rm ln}\left( R_{\rm s}+r \right)}{r} - \frac{{\rm ln}R_{\rm s}}{r} \right) dr \\
& = 16 \pi^2 {\rm G} \rho_0 \rho_{\rm hot} R_{\rm s}^3 \int_0^{R_{\rm vir}} r \left( {\rm ln}\left(R_{\rm s}+r\right) - {\rm ln}R_{\rm s} \right) dr \\
% & = 16 \pi^2 {\rm G} \rho_0 \rho_{\rm hot} R_{\rm s}^3 \\
% & \left\{\frac{1}{4} \left[ r\left(2 R_{\rm s} - r\right) -2\left(R_{\rm s}^2 - r^2\right) {\rm ln}\left(R_{\rm s} + r\right) \right] -\frac{1}{2} r^2 {\rm ln}\left(R_{\rm s}\right) \right\} \Bigg\rvert^{R_{\rm vir}}_0 \\
% & = 16 \pi^2 {\rm G} \rho_0 \rho_{\rm hot} R_{\rm s}^3 R_{\rm vir}^2 \times \\
% & \left[ \frac{1}{2} \left( \frac{R_{\rm s}}{R_{\rm vir}}\right) -\frac{1}{4} -\frac{1}{2}\frac{R_{\rm s}^2}{R_{\rm vir}^2}{\rm ln}\left(\frac{R_{\rm s}+R_{\rm vir}}{R_{\rm s}}\right) +\frac{1}{2} {\rm ln}\frac{\left(R_{\rm s}+R_{\rm vir}\right)}{R_{\rm s}} \right] \\
% & = 16 \pi^2 {\rm G} \frac{\mvir} {4 \pi R_{\rm s}^3 \left[ \frac{-R_{\rm vir}}{R_{\rm s} + R_{\rm vir}} + {\rm ln}\frac{R_{\rm s} + R_{\rm vir}}{R_{\rm s}} \right]} \frac{3 M_{\rm hot}} {4 \pi R_{\rm vir}^3} R_{\rm s}^3 R_{\rm vir}^2 \times \\
% & \left[ \frac{1}{2} \left(1-\frac{R_{\rm s}^2}{R_{\rm vir}^2}\right) {\rm ln}\left(\frac{R_{\rm s}+R_{\rm vir}}{R_{\rm s}}\right) +\frac{1}{2} \left( \frac{R_{\rm s}}{R_{\rm vir}}\right) -\frac{1}{4} \right] \\
& = \frac{3 {\rm G} \mvir M_{\rm hot}} {R_{\rm vir} \left[ \frac{-R_{\rm vir}}{R_{\rm s} + R_{\rm vir}} + {\rm ln}\frac{R_{\rm s} + R_{\rm vir}}{R_{\rm s}}\right]} \times \\
& \left[ \frac{1}{2} \left(1-\frac{R_{\rm s}^2}{R_{\rm vir}^2}\right) {\rm ln}\left(\frac{R_{\rm s}+R_{\rm vir}}{R_{\rm s}}\right) +\frac{1}{2} \left( \frac{R_{\rm s}}{R_{\rm vir}}\right) -\frac{1}{4} \right] \\
\end{aligned}
\end{equation}

Similarly, the total binding energy of cold gas can be decomposed into four components:
\begin{equation}
\label{eq:BEColdTotal}
\begin{aligned}
  & E_{\rm bind, cold} = \int_{0}^{R_{\rm s}} dM_{\rm cold} U(r) \\
  & = E_{\rm cold,DM} + E_{\rm cold,stellar} + E_{\rm cold, cold} + E_{\rm cold,hot} \\
\end{aligned}
\end{equation}

\begin{equation}
\label{eq:BEColdStellar}
\begin{aligned}
& E_{\rm cold,stellar} = \int_0^{R_{\rm s}} dM_{\rm cold} \int_r^{\infty}\frac{{\rm G}\mste(R)}{{R}^2} dR \\
& = \int_0^{R_{\rm s}} dM_{\rm cold} \left[ \int_r^{R_{\rm s}} \frac{{\rm G} \mste(R)}{{R}^2} dR + \int_{R_{\rm s}}^{\infty} \frac{{\rm G}\mste(R)}{{R}^2} dR \right] \\
% & = \int_0^{R_{\rm s}} dM_{\rm cold} \left[ \int_r^{R_{\rm s}} \frac{{\rm G} 4 \pi \rho_{\rm star} R}{3} dR + \int_{R_{\rm s}}^{\infty} \frac{{\rm G}\mste}{{R}^2} dR \right] \\
% & = \int_0^{R_{\rm s}} dM_{\rm cold} \left[ \frac{{\rm G} 2 \pi \rho_{\rm star} R^2}{3} \Bigg\rvert_r^{R_{\rm s}} + \frac{-{\rm G}\mste}{{R}} \Bigg\rvert_{R_{\rm s}}^{\infty} \right] \\
% & = \int_0^{R_{\rm s}} dM_{\rm cold} \left[ \frac{{\rm G} 2 \pi \rho_{\rm star} (R_{\rm s}^2 - r^2) }{3}  + \frac{{\rm G}\mste}{R_{\rm s}} \right] \\
& = \int_0^{R_{\rm s}} 4 \pi r^2 \rho_{\rm cold} dr \left[ \frac{{\rm G} 2 \pi \rho_{\rm star} (R_{\rm s}^2 - r^2) }{3}  + \frac{{\rm G}\mste}{R_{\rm s}} \right] \\
% & = 4 \pi \rho_{\rm cold} \left[ \frac{{\rm G} 2 \pi \rho_{\rm star} (\frac{r^3}{3}R_{\rm s}^2 - \frac{r^5}{5}) }{3}  + \frac{{\rm G}\mste r^3}{3R_{\rm s}} \right] \Bigg\rvert_0^{R_{\rm s}} \\
& = 4 \pi \rho_{\rm cold} \left[ \frac{{\rm G} 2 \pi \rho_{\rm star} (\dfrac{{R_{\rm s}}^3}{3}R_{\rm s}^2 - \dfrac{{R_{\rm s}}^5}{5}) }{3}  + \frac{{\rm G}\mste{R_{\rm s}}^3}{3R_{\rm s}} \right] \\
& = \frac{6{\rm G} M_{\rm cold} \mste}{5R_{\rm s}}
\end{aligned}
\end{equation}

\begin{equation}
\label{eq:BEColdHot}
\begin{aligned}
& E_{\rm cold,hot} = \int_0^{R_{\rm s}} dM_{\rm cold} \int_r^{\infty}\frac{{\rm G}M_{\rm hot}(R)}{{R}^2} dR \\
& = \int_0^{R_{\rm s}} dM_{\rm cold} \left[ \int_r^{R_{\rm vir}}\frac{{\rm G}M_{\rm hot}(R)}{{R}^2} dR + \int_{R_{\rm vir}}^\infty \frac{{\rm G}M_{\rm hot}(R)}{{R}^2} dR \right]\\
% & = \int_0^{R_{\rm s}} dM_{\rm cold} \left[ \int_r^{R_{\rm vir}}\frac{{\rm G} \rho_{\rm hot} 4 \pi R}{3} dR + \int_{R_{\rm vir}}^\infty \frac{{\rm G}M_{\rm hot}}{{R}^2} dR \right]\\
% & = \int_0^{R_{\rm s}} dM_{\rm cold} \left[ \frac{{\rm G} \rho_{\rm hot} 2 \pi R^2}{3} \Bigg\rvert_r^{R_{\rm vir}} +  \frac{-{\rm G}M_{\rm hot}}{R} \Bigg\rvert_{R_{\rm vir}}^\infty \right]\\
% & = \int_0^{R_{\rm s}} dM_{\rm cold} \left[ \frac{{\rm G} \rho_{\rm hot} 2 \pi (R_{\rm vir}^2 - r^2)}{3} + \frac{{\rm G}M_{\rm hot}}{R_{\rm vir}} \right]\\
& = \int_0^{R_{\rm s}} \rho_{\rm cold} 4 \pi r^2 dr \left[ \frac{{\rm G} \rho_{\rm hot} 2 \pi (R_{\rm vir}^2 - r^2)}{3} + \frac{{\rm G}M_{\rm hot}}{R_{\rm vir}} \right]\\
% & = \rho_{\rm cold} 4 \pi \left[ \frac{{\rm G} \rho_{\rm hot} 2 \pi (\frac{r^3}{3} R_{\rm vir}^2 - \frac{r^5}{5})}{3} + \frac{{\rm G}M_{\rm hot} r^3}{3R_{\rm vir}} \right] \Bigg\rvert_0^{R_{\rm s}}\\
& = \rho_{\rm cold} 4 \pi \left[ \frac{{\rm G} \rho_{\rm hot} 2 \pi (\dfrac{R_{\rm s}^3}{3} R_{\rm vir}^2 - \dfrac{R_{\rm s}^5}{5})}{3} + \frac{{\rm G}M_{\rm hot} R_{\rm s}^3}{3R_{\rm vir}} \right] \\
& = \frac{{\rm G} M_{\rm hot} M_{\rm cold}}{R_{\rm vir}}\left(\frac{3}{2}-\frac{3R_{\rm s}^2 }{10R_{\rm vir}^2 }\right)
\end{aligned}
\end{equation}

\begin{equation}
\label{eq:BEColdDM} 
\begin{aligned}
& E_{\rm cold,DM} = \int_0^{R_{\rm s}} dM_{\rm cold} \int_r^{\infty} \frac{{\rm G}M_{\rm DM}(R)}{{R}^2} dR \\
& = 16 \pi^2 {\rm G} \rho_0 \rho_{\rm cold} R_{\rm s}^3 \int_0^{R_{\rm s}} r \left( {\rm ln}\left(R_{\rm s}+r\right) - {\rm ln}R_{\rm s} \right) dr \\
& = \frac{3 {\rm G} \mvir M_{\rm cold}}{4R_{\rm s}\left[\dfrac{-R_{\rm vir}}{R_{\rm s} + R_{\rm vir}} + {\rm ln}\dfrac{R_{\rm s} + R_{\rm vir}}{R_{\rm s}}\right]}
\end{aligned}
\end{equation}

Finally, $E_{\rm cold, cold}$ can be estimated by
\begin{equation}
\label{eq:BEColdCold}
\begin{aligned}
E_{\rm cold,cold} = \frac{3{\rm G}M_{\rm cold}^2}{5R_{\rm s}}
\end{aligned}
\end{equation}

\section{Clustered vs. individual HII regions}
\label{appendix:cihii}
\begin{figure}
  \centering
  \includegraphics[width=\columnwidth]{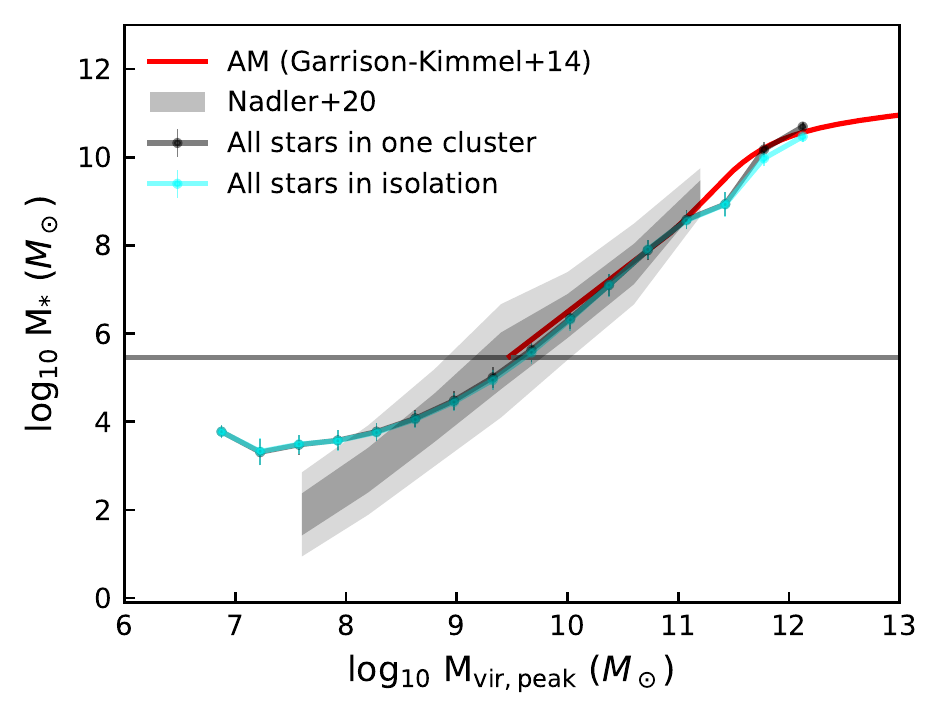}
  \caption{ We show the SMHM relation from 30 \ctp trees if we assume that all massive stars reside in one big cluster in black and that all massive stars form in isolation in cyan. }
  \label{fig:clustindi}
\end{figure}

As derived in Appendix \ref{appendix:MCR}, $\dot{M}_{\rm heat}$ does not depend linearly on the number of ionizing photons. Therefore, whether stars are in clusters or isolation is important to the stellar feedback. We show two extreme cases of star clustering: 1) all stars reside in the very central region and can be considered as one cluster, and 2) all stars are in isolation. The resulting SMHM relations of these two cases are shown in Figure~\ref{fig:clustindi}. The distinct difference lies in the most massive haloes, i.e., the MW in each \ctp tree. In such haloes, the number of massive stars is bigger and therefore the importance of $\dot{M}_{\rm heat}$ emerges.

\section{Derivation of $M_0$ in model F13}
\label{appendix:m0}
%%%%% detailed derivation of M_0 
We show the derivation of Eq.~\ref{eq:M_0} here.
Following \citet{Bryan:1998aa}, the mean density of a halo can be written as
\begin{equation}
    \overline{\rho}_{\rm h} = \frac{3 M_{\rm vir}}{4 \pi r_{\rm vir}^3} = \Delta_{\rm vir} \rho_{m}(z) = \Delta_{\rm vir} \Omega_{\rm m}(z) \frac{3 H^2(z)}{8 \pi G},
\end{equation}
where $M_{\rm vir}$ is the virial mass of the halo, $\Delta_{\rm vir} = \overline{\rho}_{\rm h} / \rho_{m}(z)$ is the over density, $\Omega_{\rm m}(z)$ is the redshift dependent matter density, $H(z)$ is the Hubble constant, and $G$ is the gravitational constant. We have the following equations: 
\begin{equation}
    \rho_{\rm m} (z) = \Omega_{\rm m} (z) \rho_{\rm c} (z) = \rho_{\rm m, 0} (1+z)^{3} = \Omega_{\rm m,0} \frac{3 H^2_{0}}{8 \pi G} (1+z)^3, 
\end{equation}
and 
\begin{equation}
\label{eq:rvir_Mvir}
    r_{\rm vir}^3 = h^{-2} M_{\rm vir} \Delta_{\rm vir}^{-1} \Omega_{\rm m,0}^{-1} (1+z)^{-3} \frac{3}{4 \pi} \frac{8 \pi G}{3 H^2_{0}}, 
\end{equation}
where $H_{\rm 0} = 67.8 ~\rm {km s^{-1} ~Mpc^{-1}}$ \citep{Planck15XIII}.
We can rewrite Eq.~\ref{eq:rvir_Mvir} into 
\begin{equation}
\begin{aligned}
    r_{\rm vir} \simeq & 210 h^{-1} \rm {kpc} \left( \frac{M_{\rm vir}}{10^{12} h^{-1}\msun} \right) ^{1/3} \\
    & \times \left( \frac{\Delta_{\rm vir}}{200} \right) ^{-1/3} \Omega_{\rm m,0}^{-1/3} (1+z)^{-1}
\end{aligned}
\end{equation}
Therefore, the circular velocity at virial radius is 
\begin{equation}
\begin{aligned}
    & v_{\rm cir} = \sqrt{\frac{G M_{\rm vir}}{r_{\rm vir}}} \simeq 146.6 \rm{km s^{-1}} \\ 
    &\left(\frac{M_{\rm vir}}{10^{12} \rm{h^-1} \msun} \right)^{1/3} \left( \frac{\Delta_{\rm vir}}{200} \right)^{1/6} \Omega_{\rm m,0}^{1/6} (1+z)^{1/2} 
\end{aligned}
\end{equation}

Finally, we have $M_{\rm vir}$ as a function of $v_{\rm cir}$, over density $\Delta_{\rm vir}$, and redshift $z$, 
\begin{equation}
\begin{aligned}
    & \frac{M_{\rm vir}}{10^{12} \rm{h}^{-1} \msun} = \\
    & \left( \frac{v_{\rm cir} }{146.6 \rm{km s^{-1}} } \right)^3 \left( \frac{\Delta_{\rm vir}}{200} \right)^{-1/2} \Omega_{\rm m,0}^{-1/2} (1+z)^{-3/2}. 
\end{aligned}
\end{equation}

Here we assume that $M_{\rm vir}$ is at $\Delta_{\rm vir}=200$ and substitute $M_{\rm vir} = M_0$ and $v_{\rm cir} = v_{\rm cool}$ to arrive at Eq.~\ref{eq:M_0}.
%%%%% detailed derivation of M_0 

%%%%%%%%%%%%%%%%%%%%%%%%%%%%%%%%%%%%%%%%%%%%%%%%%%

% Don't change these lines
\bsp	% typesetting comment
\label{lastpage}
\end{document}